\documentclass[a4paper,12pt]{article}

\usepackage{cite}
\usepackage{epsfig}
\usepackage{amssymb,amsmath}
\usepackage{mathrsfs}
\usepackage{graphicx,subfigure}

\pdfoutput=1

\newlength{\dinwidth}
\newlength{\dinmargin}
\newcommand{\Br}{\mathrm{Br}}

\newcommand{\cP}{\mathcal{P}}

\def\be{\begin{equation}}
\def\ee{\end{equation}}
\def\beqn{\begin{eqnarray}}
\def\eeqn{\end{eqnarray}}

\def\ba{\begin{array}{c}}
\def\bat{\begin{array}{cc}}
\def\ea{\end{array}}
\def\bi{\begin{itemize}}
\def\ei{\end{itemize}}

\def\cL{{\cal L}}

\usepackage{color}

\definecolor{DarkGreen}{rgb}{0.65,1,0.75}
\definecolor{violet}{rgb}{0.5,0,1}

\begin{document}

\title{
\begin{flushright}\vbox{\normalsize \mbox{}\vskip -6cm FTUV/12$-$1031 \\[-3pt] IFIC/12$-$59\\[-3pt] DO-TH 12/32} %%% \\[-3pt]\today}
\end{flushright}
\vskip 30pt
{\bf Sensitivity to charged scalars in \\ $\boldsymbol{B\to D^{(*)}\tau\nu_\tau}$\ and $\boldsymbol{B\to\tau\nu_\tau}$\ decays}}
\bigskip\bigskip

\author{Alejandro Celis$^{1}$, Martin Jung$^{2}$, Xin-Qiang Li$^{3,1}$ and Antonio Pich$^{1}$\\[20pt]
{$^1$\small IFIC, Universitat de Val\`encia -- CSIC, Apt. Correus 22085, E-46071 Val\`encia, Spain}\\
{$^2$\small Institut f\" ur Physik, Technische Universit\" at Dortmund, D-44221 Dortmund, Germany}\\
{$^3$\small Department of Physics, Henan Normal University, Xinxiang, Henan 453007, P.~R. China}}

\date{}
\maketitle
\bigskip \bigskip

\begin{abstract}
\noindent
We analyze the recent experimental evidence for an excess of $\tau$-lepton production in several exclusive semileptonic $B$-meson decays in the context of two-Higgs-doublet models. These decay modes are sensitive to the exchange of charged scalars and constrain strongly their Yukawa interactions. While the usual Type-II scenario cannot accommodate the recent BaBar data, this is possible within more general models in which the charged-scalar couplings to up-type quarks are not as suppressed. Both the $B\to D^{(*)}\tau\nu_\tau$ and the $B\to\tau\nu_\tau$ data can be fitted within the framework of the Aligned Two-Higgs-Doublet Model, but the resulting parameter ranges are in conflict with the constraints from leptonic charm decays. This could indicate a departure from the family universality of the Yukawa couplings, beyond their characteristic fermion mass dependence. We discuss several new observables that are sensitive to a hypothetical charged-scalar contribution, demonstrating that they are well suited to distinguish between different scenarios of new physics in the scalar sector, and also between this group and models with different Dirac structures; their experimental study would therefore shed light on the relevance of scalar exchanges in semileptonic $b\to c\,\tau^-\bar\nu_\tau$ transitions.
\end{abstract}

\newpage

\section{Introduction}
\label{Sec:intro}

The BaBar collaboration has recently reported an excess of events in two semileptonic transitions of the type $b\to c\,\tau^-\bar\nu_\tau$. More specifically, they have measured the ratios~\cite{Lees:2012xj}
\beqn\label{eq:Babar}
R(D)& \equiv &\frac{\Br(\bar B\to D\tau^-\bar\nu_\tau)}{\Br(\bar B\to D\ell^-\bar\nu_\ell)}\; \stackrel{\rm BaBar}{=}\; 0.440\pm 0.058\pm 0.042\;\stackrel{\rm avg}{=}\; 0.438\pm0.056\, ,
\nonumber\\[5pt]
R(D^*)& \equiv &\frac{\Br(\bar B\to D^*\tau^-\bar\nu_\tau)}{\Br(\bar B\to D^*\ell^-\bar\nu_\ell)}\; \stackrel{\rm BaBar}{=}\; 0.332\pm 0.024\pm 0.018\;\stackrel{\rm avg}{=}\; 0.354\pm0.026\, ,
\eeqn
which are normalized to the corresponding decays into light leptons $\ell=e,\mu$. The second value given in each line is the average with the previous measurements by the Belle collaboration~\cite{Adachi:2009qg,Bozek:2010xy}, which also yield central values corresponding to an excess, but not significantly so. These relative rates can be predicted with a rather high accuracy, because many hadronic uncertainties cancel to a large extent. The Standard Model~(SM) expectations~\cite{Fajfer:2012jt,Sakaki:2012ft,Fajfer:2012vx,Becirevic:2012jf} are significantly lower than the BaBar measurements. If confirmed, this could signal new physics~(NP) contributions violating lepton-flavour universality.

A sizable deviation with respect to the SM prediction was previously observed in the leptonic decay $B^-\to\tau^-\bar\nu_\tau$, when combining the data by BaBar~\cite{Aubert:2009wt,Lees:2012ju} and Belle~\cite{Hara:2010dk}. The world average $\Br(B^-\to\tau^-\bar\nu_\tau) = (1.65\pm 0.34)\times 10^{-4}$~\cite{Beringer:1900zz} used to be $2.5\sigma$ higher than the SM prediction $(0.733\,{}^{+\,0.121}_{-\,0.073})\times 10^{-4}$~\cite{Charles:2004jd} (taking the modulus of the Cabibbo-Kobayashi-Maskawa~(CKM)~\cite{CKM} matrix element $|V_{ub}|$ from a global CKM fit). However, a value closer to the SM expectation has been just reported by Belle~\cite{Adachi:2012he}, leading to the new Belle combination $\Br(B^-\to\tau^-\bar\nu_\tau)_{\mathrm{Belle}} = (0.96\pm 0.22\pm 0.13)\times 10^{-4}$, which we average with the combined BaBar result~\cite{Lees:2012ju} to obtain $\Br(B^-\to\tau^-\bar\nu_\tau) = (1.15\pm 0.23)\times 10^{-4}$.

While more experimental studies are clearly needed, these measurements are intriguing enough to trigger the theoretical interest~\cite{Fajfer:2012jt,Fajfer:2012vx,Sakaki:2012ft,Becirevic:2012jf,
Crivellin:2012ye,Datta:2012qk,Bailey:2012jg,Faustov:2012nk,Deshpande:2012he,Choudhury:2012hn}. This kind of non-universal enhancement of the $\tau$ production in semileptonic $B$-meson decays could be generated by NP contributions with couplings proportional to fermion masses. In particular, it could be associated with the exchange of a charged scalar within two-Higgs-doublet models~(2HDMs), with the expected contribution to the transition amplitude being proportional to $m_\tau m_b/M_{H^\pm}^2$. This approach offers (obviously) a solution when considering scalar NP contributions model-independently; even in this general case, however, non-trivial predictions for other observables in these decays can be obtained. More specific models generally have difficulties in describing all the data. For example, the BaBar data on $\bar B \to D\,\tau^-\bar\nu_\tau$ and $\bar B \to D^{*}\tau^-\bar\nu_\tau$ cannot be explained simultaneously within the usually adopted Type-II scenario~\cite{Lees:2012xj,Fajfer:2012jt,Crivellin:2012ye}. It is also observed that none of the four types of 2HDMs with natural flavour conservation~(i.e., Type-I, Type-II, ``lepton specific" and ``flipped")~\cite{2HDM-review} can simultaneously account for the $B\to \tau$ data~\cite{Fajfer:2012jt}. We shall show, however, that they can be accommodated by the more general framework of the Aligned Two-Higgs-Doublet Model~(A2HDM)~\cite{Pich:2009sp}, albeit creating a tension when including them in a global fit.

Other suggested interpretations of the observed excess within different NP scenarios include the 2HDM of Type-III, equipped with a MSSM-like Higgs potential and flavour-violation in the up-quark sector~\cite{Crivellin:2012ye}, a leptoquark model with renormalizable interactions to third-generation SM fermions~\cite{Fajfer:2012jt}, composite Higgs models where the heavier SM fermions are expected to be partially or mostly composite~\cite{Fajfer:2012jt}, the exchange of right-handed down-type squarks within the R-parity violating MSSM~\cite{Deshpande:2012he}, as well as a non-universal left-right model where only the third generation couples to the $W_R$~\cite{He:2012zp}.

Generic multi-Higgs-doublet models give rise to unwanted flavour-changing neutral current~(FCNC) interactions through non-diagonal couplings of neutral scalars to fermions~\cite{2HDM-review}. The tree-level FCNCs can be eliminated by requiring the alignment in flavour space of the Yukawa matrices coupling to a given right-handed fermion~\cite{Pich:2010ic}. This results in a very specific structure, with all fermion-scalar interactions being proportional to the corresponding fermion masses, and implies an interesting hierarchy of FCNC effects, suppressing them in light-quark systems while allowing potentially relevant signals in heavy-quark transitions.
The A2HDM leads to a rich and viable phenomenology~\cite{Pich:2009sp,Pich:2010ic,Jung:2010ik,Jung:2010ab,Jung:2012vu}; it constitutes a very general framework which includes, for particular values of its parameters, all previously considered 2HDMs without tree-level FCNCs~\cite{2HDM-review}, and at the same time incorporates additional new sources of CP violation beyond the SM.

In the following, we shall consider the phenomenology of $b\to q\,\tau^-\bar\nu_\tau$ ($q=u,c$) transitions within a framework with additional scalar operators, assumed to be generated by the exchange of a charged scalar. Starting from the most general parametrization of such effects, we then specialize to various more specific models to examine their capability to describe the data and the possibility to distinguish between them.

Our paper is organized as follows: In Sec.~\ref{sec:theory}, we briefly describe the theoretical framework adopted in our analysis. In Sec.~\ref{sec:discussion}, we present our numerical results and show the parameter ranges needed to explain the present data. We proceed in Sec.~\ref{sec:observables} by analyzing various additional observables sensitive to scalar contributions, both integrated and differential, before concluding in Sec.~\ref{sec:conclusion}. The appendices include a discussion of the relevant input parameters and details on the calculation for the semileptonic $B$-meson decays.

\section{Theoretical framework}
\label{sec:theory}

We are going to assume that, in addition to the SM $W$-exchange amplitude, the quark-level transitions $b\to q\,l^-\bar\nu_l$ receive tree-level contributions from the exchange of a charged scalar. The effective low-energy Lagrangian describing these transitions takes then the form
\be\label{genlagrangian}
\cL_{\rm eff} \; =\; -\frac{4 G_F}{\sqrt{2}}\;\sum_{q=u,c}\, V_{qb}\;\sum_{l=e,\mu,\tau}\;\left\{
\left[\bar q\gamma^\mu \cP_L b\right] \left[\bar l\gamma_\mu\cP_L\nu_l\right]
\, +\, \left[\bar q \left( g_L^{qbl}\,\cP_L + g_R^{qbl}\, \cP_R\right) b\right] \left[\bar l\cP_L\nu_l\right]
\right\}\, ,
\ee
where $\mathcal P_{R,L} \equiv \frac{1\pm \gamma_5}{2}$ are the chirality projectors, and the effective couplings are, in the majority of 2HDMs, proportional to the fermion masses, $g_L^{{q_u}{q_d}l}\sim m_{q_u}m_l/M_{H^\pm}^2$, $g_R^{{q_u}{q_d}l}\sim m_{q_d}m_l/M_{H^\pm}^2$. These explicit fermion mass factors imply negligible effects in decays into light leptons~($e$, $\mu$),\footnote{
%%%%%%%
The obvious exception are leptonic meson decays, where the SM contribution is already suppressed by the light lepton masses, yielding a large relative contribution from the charged scalar.}
%%%%%%%
while decays involving the $\tau$ receive potentially large contributions. Owing to the $m_q$ suppression, the coupling $g_L^{qbl}$ does not play any relevant role in $b\to u \tau^-\bar\nu_\tau$ transitions, but it can give sizeable corrections to $b\to c \tau^-\bar\nu_\tau$.

We present next the key relations for including scalar NP contributions in the decays in question. When considering specific models, the main focus lies on the A2HDM, of which we give a short review. For the relevant kinematical variables, notation and derivation of the double differential decay rates, we refer the reader to the appendices.

\subsection{$\boldsymbol{b\to q\,\boldsymbol{\tau^-\,\bar\nu_\tau}~(q=u,c)}$ decays}
\label{sec:b-decays}

Due to the helicity suppression of the SM amplitude, the leptonic decay $B^-\to\tau^-\bar\nu_\tau$ is particularly sensitive to the charged scalar exchange. The total decay width is given by~\cite{Jung:2010ik,Hou:1992sy}
\begin{equation}\label{eq:Gamma_Plnu}
\Gamma(B^-\to\tau^-\bar\nu_\tau)\, =\, G_F^2m_\tau^2f_B^2|V_{ub}|^2 \,\frac{m_{B}}{8\pi} \left( 1- \frac{m_\tau^2}{m_{B}^2} \right)^2  (1+\delta_{\mathrm{em}})\; |1-\Delta^\tau_{ub}|^2 \; ,
\end{equation}	
where $\delta_{\mathrm{em}}$ denotes the electromagnetic radiative contributions, and the new-physics information is encoded in the correction\footnote{Here and in the following we suppress in the notation the scale dependence of {\it e.g.} the quark masses and scalar couplings.}
\begin{equation}\label{eq::Delta}
\Delta_{qb}^l\, = \,\frac{(g_L^{qbl}-g_R^{qbl}) m_B^2}{m_{l}(\overline{m}_{b} + \overline{m}_q)}\; \stackrel{q=u}{\simeq}\; -\frac{g_R^{ubl}\,m_B^2}{m_{l}\,\overline{m}_{b}}\,,
\end{equation}
absorbing in addition mass factors from the hadronic matrix elements.

Semileptonic decays receive contributions from a charged scalar as well, but in this case the leading SM amplitude is not helicity suppressed; therefore, the relative influence is smaller. In addition, they involve momentum-dependent form factors. The $\bar B\to D l\bar\nu_l$ decay amplitude is characterized by two form factors, $f_+(q^2)$ and $f_0(q^2)$, associated with the P-wave and S-wave projections of the crossed-channel matrix element $\langle 0|\bar c\gamma^\mu b|\bar B \bar D\rangle$. The scalar-exchange amplitude only contributes to the scalar form factor; it amounts to a multiplicative correction~\cite{Jung:2010ik}
\begin{equation}\label{eq:sFFmod}
f_0(q^2)\, \to \, \tilde f_0(q^2)\, = \,f_0(q^2)\;\left[ 1 + \delta_{cb}^l\,\frac{q^2}{(m_B-m_D)^2}\, \right]\, ,
\end{equation}
with
\begin{equation}  \label{eq:ww}
\delta_{cb}^l\,\equiv\, \frac{(g_L^{cbl}+g_R^{cbl})(m_B-m_D)^2}{m_l(\overline{m}_b-\overline{m}_c)}\, .
\end{equation}
The sensitivity to the scalar contribution can only be achieved in semileptonic decays into heavier leptons. The decays involving light leptons can, therefore, be used to extract information on the vector form factor, reducing the necessary theory input to information on the scalar form factor. Since the observables are usually normalized to the decays into light leptons, the relevant input quantity is not the scalar form factor itself, but the ratio of scalar to vector form factors, $f_0(q^2)/f_+(q^2)$; an important constraint on the latter is its normalization to unity at $q^2=0$. These features lead us to parametrize the different CP-conserving observables that we are going to consider in the following form:
\begin{equation}\label{eq::NPexpansion}
O\; =\; c_0\, +\, c_1\,{\rm Re}\left(\delta_{cb}^\tau\right)\, +\, c_2\left|\delta_{cb}^\tau\right|^2\,,
\end{equation}
implying a discrete symmetry ${\rm Im}\left(\delta_{cb}^\tau\right)\to -{\rm Im}\left(\delta_{cb}^\tau\right)$. The coefficients $c_i$, which contain the dependence on the strong-interaction dynamics, are parametrized in turn in terms of the vector form-factor slope $\rho_1^2$ and the scalar density $\Delta(v_B\cdot v_D)$~\cite{Kamenik:2008tj,Deschamps:2009rh}. For the former we use the value extracted from $B\to D\ell\nu$~($\ell=e,\mu$) decays. The function $\Delta(v_B\cdot v_D)\propto f_0(q^2)/f_+(q^2)$ has been studied on the lattice, in the range $v_B\cdot v_D = 1$--$1.2$, and found to be consistent with a constant value $\Delta = 0.46\pm 0.02$, very close to its static-limit approximation $(m_B-m_D)/(m_B+m_D)$~\cite{Becirevic:2012jf,Bailey:2012jg,deDivitiis:2007uk}. This value is furthermore in agreement with QCD sum rule estimates~\cite{Azizi:2008tt,Faller:2008tr}.

The decay $\bar B\to D^*l^-\bar\nu_l$ has a much richer dynamical structure, due to the vector nature of the final $D^*$ meson. The differential decay distribution is described in terms of four helicity amplitudes, $H_{\pm\pm}$, $H_{00}$ and $H_{0t}$, where the first subindex denotes the $D^*$ helicity ($\pm, 0$) and the second the lepton-pair helicity ($\pm, 0, t$) in the $B$-meson rest frame~\cite{Korner:1989qb,Hagiwara:1989gza}. In addition to the three polarizations orthogonal to its total four-momentum $q^\mu$, the leptonic system has a spin-zero time component~($t$) that is proportional to $q^\mu$ and can only contribute to the semileptonic decays for non-zero charged-lepton masses (it involves a positive helicity for the $l^-$). The corresponding $H_{0t}$ amplitude is the only one receiving contributions from the scalar exchange~\cite{Fajfer:2012vx}:
\be \label{eq:sVmode}
H_{0t}(q^2)\; =\; H_{0t}^{\mathrm{SM}}(q^2)\;\left(1 - \Delta_{cb}^\tau\, \frac{q^2}{m_B^2}\right)\,.
\ee
The observables are then given in an expansion analogous to Eq.~\eqref{eq::NPexpansion}, with $\delta_{cb}^\tau$ replaced by $\Delta_{cb}^\tau$, and the coefficients depend on the different form-factor normalizations $R_i(1)$~($i=0,1,2,3$) and the slope $\rho^2$, see Appendix~\ref{subsec:BDstar} for details. Here, again, we use inputs extracted from the decays involving light leptons where possible, while for the remaining form-factor normalization $R_{3}(1)$ we adopt a value calculated in the framework of Heavy Quark Effective Theory~(HQET)~\cite{HQET-forR31}.

A summary of the different form-factor parameters is given in Table~\ref{tab::hadronic}, in the appendix.

\subsection{Overview of the A2HDM}
\label{Sec:A2HDM-overview}

The 2HDMs extend the SM Higgs sector by a second scalar doublet of hypercharge $Y=\frac{1}{2}$. Thus, in addition to the three Goldstone bosons, they contain five physical scalars: two charged fields $H^{\pm}$ and three neutral ones $\varphi_i^0=\{h,H,A\}$. The most generic Yukawa Lagrangian with the SM fermionic content gives rise to tree-level FCNCs, because the Yukawa couplings of the two scalar doublets to fermions cannot be simultaneously diagonalized in flavour space. The non-diagonal neutral couplings can be eliminated by requiring the alignment in flavour space of the Yukawa matrices~\cite{Pich:2009sp}; {\it i.e.}, the two Yukawa matrices coupling to a given type of right-handed fermions are assumed to be proportional to each other and can, therefore, be diagonalized simultaneously. The three proportionality parameters are arbitrary complex numbers and introduce new sources of CP violation.

In terms of the fermion mass-eigenstate fields, the Yukawa interactions of the charged scalar in the A2HDM read~\cite{Pich:2009sp}
\begin{equation}\label{lagrangian}
 \mathcal L_Y^{H^\pm} =  - \frac{\sqrt{2}}{v}\, H^+ \left\{ \bar{u} \left[ \varsigma_d\, V M_d \mathcal P_R - \varsigma_u\, M_u V \mathcal P_L \right] d\, + \, \varsigma_l\, \bar{\nu} M_l \mathcal P_R l \right\}
 \,+\,\mathrm{h.c.} \,,
\end{equation}
where $\varsigma_f$~($f=u,d,l$) are the proportionality parameters in the so-called ``Higgs basis'' in which only one scalar doublet acquires a non-zero vacuum expectation value. The CKM quark mixing matrix $V$~\cite{CKM} remains the only source of flavour-changing interactions. All possible freedom allowed by the alignment conditions is determined by the three family-universal complex parameters $\varsigma_f$, which provide new sources of CP violation without tree-level FCNCs~\cite{Pich:2009sp}. Comparing Eqs.~(\ref{lagrangian}) and (\ref{genlagrangian}), one obtains the following relations between the A2HDM and the general scalar NP parameters:
\be \label{eq:hy}
g_L^{q_{u}q_{d}l}\; =\; \varsigma_u^{\phantom{*}}\varsigma_l^*\; \frac{m_{q_u} m_l}{M_{H^\pm}^2}\, ,
\qquad\qquad
g_R^{q_{u}q_{d}l}\; =\; -\varsigma_d^{\phantom{*}}\varsigma_l^*\; \frac{m_{q_d} m_l}{M_{H^\pm}^2}\, .
\ee
The usual models with natural flavour conservation~(NFC), based on discrete ${\cal Z}_2$ symmetries, are recovered for particular (real) values of the couplings $\varsigma_f$; especially $\varsigma_d = \varsigma_l = -1/\varsigma_u = - \tan \beta$ and $\varsigma_u = \varsigma_d = \varsigma_l = \cot \beta$ correspond to the Type-II and Type-I models, respectively.

Limits on the charged-scalar mass from flavour observables and direct searches depend strongly on the assumed Yukawa structure. The latest bound on the Type-II 2HDM charged Higgs from $\bar{B}\to X_s \gamma$ gives $M_{H^\pm} \ge 380~{\rm GeV}$ at $95\%$ confidence level~(CL)~\cite{Hermann:2012fc}. Within the A2HDM on the other hand it is still possible to have a light charged Higgs~\cite{Jung:2010ik,Jung:2010ab}. Assuming that the charged scalar $H^+$ only decays into fermions $u_i\bar d_j$ and $l^+\nu_l$, LEP established the limit $M_{H^\pm}> 78.6~{\rm GeV}$~($95\%$ CL)~\cite{:2001xy}, which is independent of the Yukawa structure. A charged Higgs produced via top-quark decays has also been searched for at the Tevatron \cite{Abulencia:2005jd,Abazov:2009aa} and the LHC \cite{Aad:2012tj,:2012cw}; these searches are, however, not readily translatable into constraints for the model parameters considered here. It should be noted that the charged-scalar mass enters only in combination with the other couplings and, therefore, its size does not affect directly our results at this level.

\section{Results and discussions}
\label{sec:discussion}

In Table~\ref{tab::SM} we summarize our predictions within the SM for the various semileptonic and leptonic decays considered in this work, using the hadronic inputs quoted in Table~\ref{tab::hadronic} (parameters that do not appear in this table are taken from~\cite{Beringer:1900zz}). The rates for leptonic $D$, $K$ and $\pi$ decays are obtained from Eq.~\eqref{eq:Gamma_Plnu} with appropriate replacements,
while the ratio of $\tau\to K/\pi \nu_{\tau}$ decay widths is given by~\cite{Jung:2010ik}
\be
\frac{\Gamma(\tau\to K\nu)}{\Gamma(\tau\to\pi\nu)} \; =\; \left(\frac{1-m_K^2/m_\tau^2}{1-m_\pi^2/m_\tau^2}\right)^2\, \left|\frac{V_{us}}{V_{ud}}\right|^2\, \left(\frac{f_K}{f_\pi}\right)^2\, (1+\delta_{\mathrm{em}}^{\tau K2/\tau\pi 2})\, \left|\frac{1-\Delta_{us}}{1-\Delta_{ud}}\right|^2\, .
\ee
One can see that, apart from $R({D})$ and $R({D^*})$, all the observables are in agreement with their SM predictions. While for the decays involving only
$D_{(s)}$, $K$, and $\pi$ mesons no large effect could be expected because of the relatively small quark masses involved, this is equally true for the influence of $g_L^{cb\tau}$. Contrary to that expectation, however, the data on $R({D})$ and $R({D^*})$ indicate a large value for this coupling. This not only renders especially models with NFC incompatible with the data, but also poses a problem in more general models.

%%%%%%%%%%%%%%%%%%%%%%%%%%%%%%%%%%%%%%%%%%%%%%%%%%%%%%%%%%%%%%%%%%%%%%%%%%
\begin{table}[thb]
\begin{center}
\caption{\label{tab::SM} \it \small Predictions within the SM for the various semileptonic and leptonic decays discussed in this work, together with their corresponding experimental values. The first uncertainty given always corresponds to the statistical uncertainty, and the second, when present, to the theoretical one.}
\vspace{0.2cm}
\doublerulesep 0.8pt \tabcolsep 0.07in
\small{
\begin{tabular}{lccc}
\hline\hline
Observable   					&  SM Prediction					& Exp. Value     & Comment \\
\hline
$R({D})$ 							& $0.296^{+0.008}_{-0.006}\pm0.015$ & $ 0.438 \pm 0.056$ & our average~\cite{Lees:2012xj,Adachi:2009qg,Bozek:2010xy} \\
$R({D^*})$       					& $0.252\pm0.002\pm0.003$ 		& $0.354 \pm 0.026$						 & our average~\cite{Lees:2012xj,Adachi:2009qg,Bozek:2010xy} \\
$\Br(B\to \tau \nu_\tau)$ 	&  $(0.79^{+0.06}_{-0.04}\pm0.08)\times 10^{-4}$& $(1.15 \pm 0.23)\times 10^{-4}$	& our average~\cite{Adachi:2012he,Lees:2012ju} \\
$\Br(D_s \to \tau \nu_\tau)$	&  $(5.18 \pm 0.08\pm0.17) \times 10^{-2} $& $(5.54 \pm 0.24)\times 10^{-2}$& our average~\cite{Beringer:1900zz,Wang} \\
$\Br(D_s \to \mu \nu)$	& $(5.31 \pm 0.09\pm0.17) \times 10^{-3} $	& $(5.54 \pm 0.24)\times 10^{-3}$	& our average~\cite{Beringer:1900zz,Wang} \\
$\Br(D \to \mu \nu)$	&  $(4.11^{+0.06}_{-0.05}\pm0.27) \times 10^{-4} $ & $(3.76 \pm 0.18)\times 10^{-4}$	 & \cite{Rong:2012at} \\
$\Gamma(K\to\mu\nu)/\Gamma(\pi\to\mu\nu)$    & $1.333\pm0.004\pm0.026$ & $1.337\pm0.003$ & \cite{Beringer:1900zz} \\
$\Gamma(\tau\to K\nu_\tau)/\Gamma(\tau\to\pi\nu_\tau)$ & $(6.56\pm0.02\pm0.15)\times10^{-2}$ & $(6.46\pm0.10)\times10^{-2}$ & \cite{Beringer:1900zz} \\
\hline\hline
\end{tabular}}
\end{center}
\end{table}
%%%%%%%%%%%%%%%%%%%%%%%%%%%%%%%%%%%%%%%%%%%%%%%%%%%%%%%%%%%%%%%%%%%%%%%%%%

We start by analyzing the constraints on the A2HDM parameters from the decays listed in Table~\ref{tab::SM}.\footnote{We do not take into account the experimental correlation between the measured values of $R(D)$ and $R(D^*)$ given in \cite{Lees:2012xj}. It is reduced to $-19\%$ when averaging with the Belle data and does not affect our results significantly. Moreover, the BaBar fit is sensitive to the assumed kinematical distribution, which is modified by the scalar contribution. While BaBar has already performed an explicit analysis within the Type-II 2HDM, it would be useful to analyze the experimental data in terms of the more general complex parameters $\Delta^l_{cb}$ and $\delta^l_{cb}$, to make the inclusion of this effect possible in the future. This modification is, however, only relevant for large values of the scalar couplings, which are excluded in the scenarios~2 and 3 discussed below.}
In contrast to the models with NFC, the observables involving $B$-meson decays~($R({D})$, $R({D^*})$ and $\Br(B\to\tau\nu)$) can be consistently explained in the A2HDM.
However, the resulting parameter region excludes the one selected by the leptonic $D_{(s)}$-meson decays. More generally speaking, models fulfilling the relations
\begin{equation} \label{eq::gratios}
(a) \quad g_L^{q_{u}q_{d}l}/g_L^{q'_{u}q'_{d}l'}=m_{q_u}m_l/(m_{q'_u}m_{l'}) \qquad {\rm and } \qquad
(b) \quad g_R^{q_{u}q_{d}l}/g_R^{q'_{u}q'_{d}l'}=m_{q_d}m_l/(m_{q'_d}m_{l'})
\end{equation}
are in conflict with the data. Removing $R({D^*})$ from the fit leads to a consistent picture in the A2HDM; in this case, however, the SM is also globally consistent with the data, as the tension in $R(D)$ is ``distributed'' over the remaining observables. Models with NFC remain disfavoured compared to the SM. These observations lead us to consider the following scenarios:
\begin{enumerate}
\item[$\bullet$] Scenario~1~(Sc.1) is a model-independent approach where all couplings $g_{L,R}^{q_{u}q_{d}l}$ are assumed to be independent. One possible realization is the 2HDM of Type III. This implies that the effective couplings $\delta_{cb}^l$ and $\Delta_{cb}^l$ in the two semileptonic processes can be regarded as independent. Therefore, predictions for the additional observables in $B\to D(D^*)\tau\nu$ follow in this case \emph{only} from $R(D)$~($R(D^*)$).

\item[$\bullet$] In scenario~2 we assume the relations in Eq.~(\ref{eq::gratios}) to hold for $q_d=b$, while processes involving only the first two generations are regarded as independent. When considering only the constraints from $R({D})$, $R({D^*})$ and $\Br(B\to\tau\nu)$, the couplings in the A2HDM fulfill this condition; we will assume this form in the following for definiteness.

\item[$\bullet$] For scenario~3 we discard the measurement of $R({D^*})$ as being due to a statistical fluctuation and/or an underestimated systematic effect, leaving us with a viable A2HDM. From a global fit to all the other measurements we then obtain predictions for the new observables in $B\to D^{(*)}\tau\nu$ as well as $R({D^*})$.
\end{enumerate}
The ratios in Eq.~\eqref{eq::gratios} might of course also be changed for $l=\tau$ or $l=\mu$. However, because of the smallness of $m_\mu$, this would not be visible in any of the observables considered here. A way to test this option is to consider the ratio $\mathrm{Br}(B^-\to\tau^-\bar\nu_\tau)/\mathrm{Br}(B^-\to\mu^-\bar\nu_\mu)$, which is independent of NP in the scalar sector if Eq.~\eqref{eq::gratios} is fulfilled.

\begin{figure}[thb]
\centering
\includegraphics[height=5.7cm]{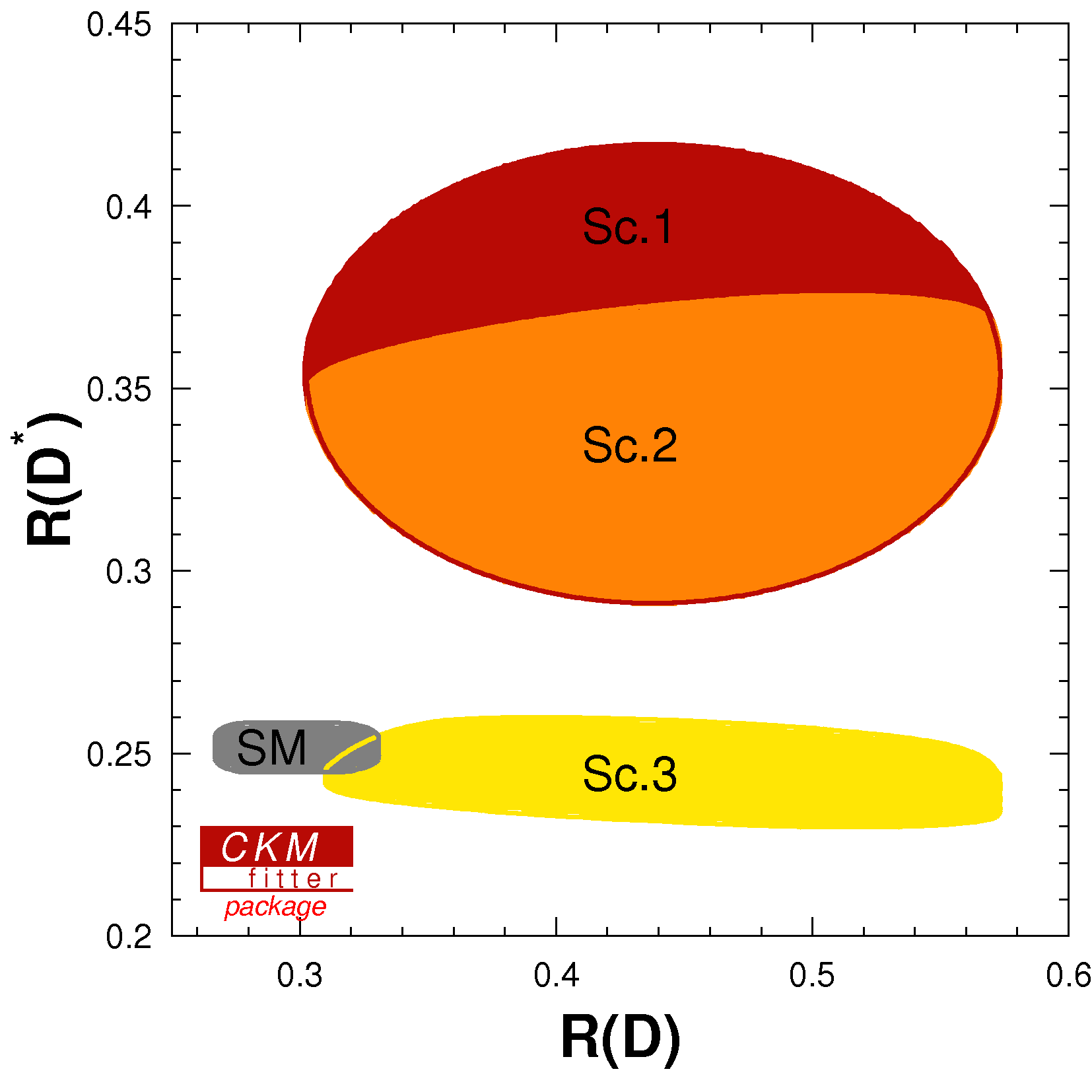}
% \hspace{0.2cm}
\includegraphics[height=5.7cm]{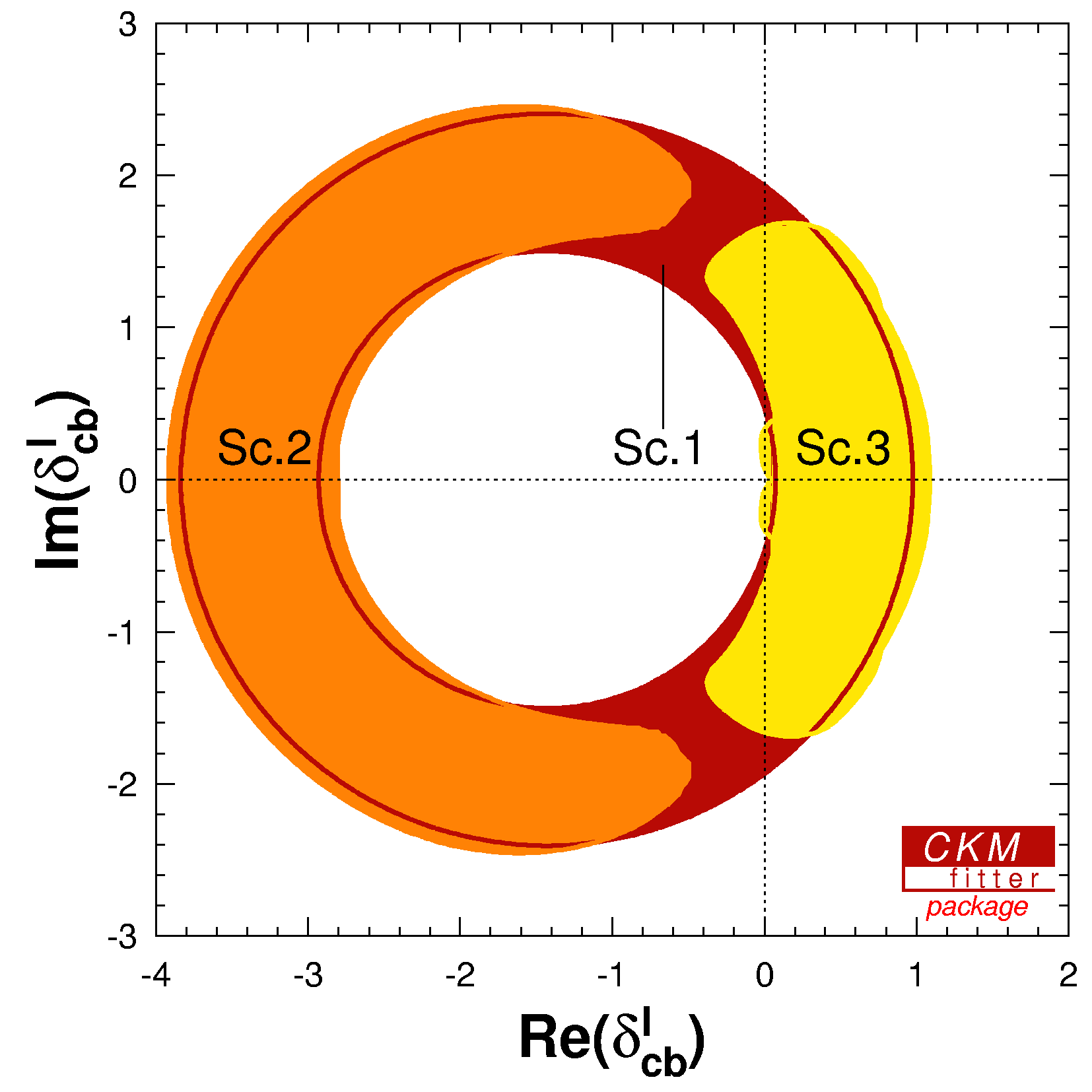}
% \hspace{0.2cm}
\includegraphics[height=5.7cm]{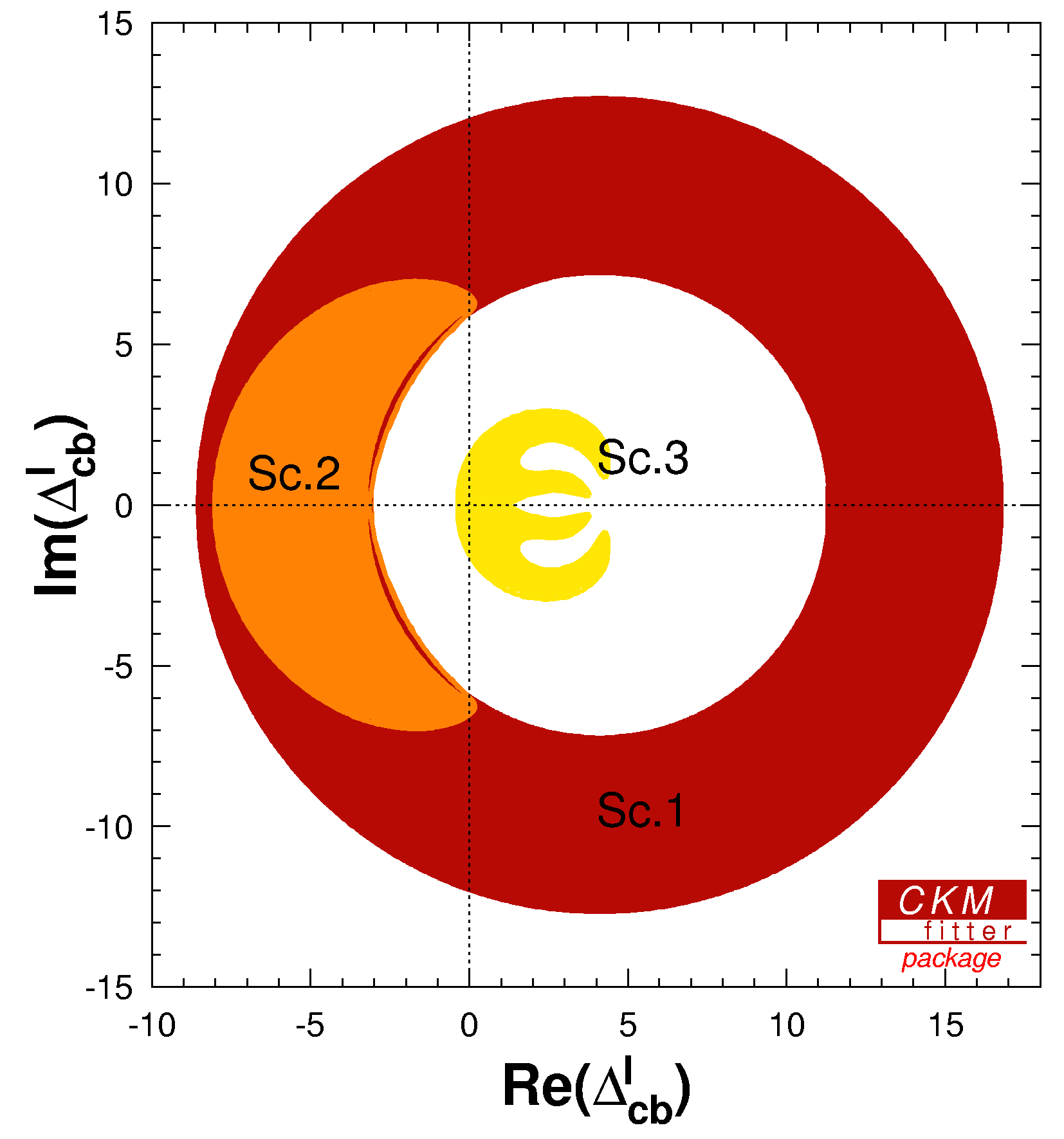}
\caption{\label{zetadlul-allowed} \it \small Allowed regions in the $R(D)$--$R(D^*)$~(left), complex $\delta_{cb}^l$~(center) and $\Delta_{cb}^l$~(right) planes at $95\%$~CL, corresponding to the three different scenarios. See text for details.}
\end{figure}

In Fig.~\ref{zetadlul-allowed} we first show the allowed regions in the $R(D)$--$R(D^*)$ plane for the different scenarios. Scenario~1 is just reflecting the experimental information, while the additional constraint from $B\to\tau\nu$ already excludes part of that area in scenario~2. For the third scenario, the tension with the measurement of $R(D^*)$ is clearly visible; the allowed range includes the SM range and values even further away from the measurement, thereby predicting this effect to vanish completely in the future if this scenario is realized. The value implied by the fit reads $R(D^*)_{\rm Sc.3}=0.241\pm0.003\pm0.007$.

%%%%%%%%%%%%%%%%%%%%%%%%%%%%%%%%%%%%%%%%%%%%%%%%%%%%%%%%%%%%%%%%%%%%%%%%%%
%%%%%%%%%%%%%%%%%%%%%%%%%%%%%%%%%%%%%%%%%%%%%%%%%%%%%%%%%%%%%%%%%%%%%%%%%%

\begin{figure}[thb]
\centering
\includegraphics[height=6.8cm]{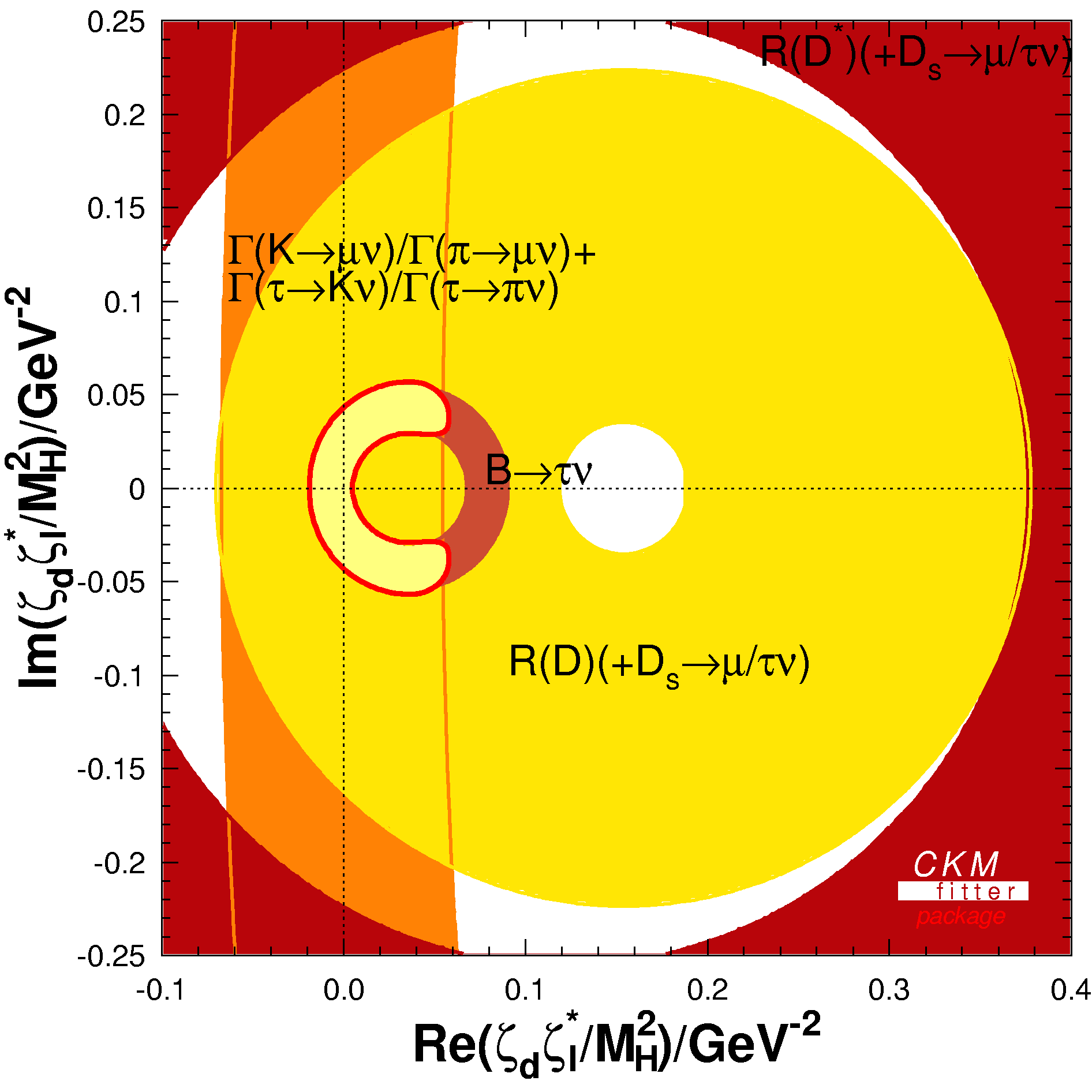}
\hspace{0.5cm}
\includegraphics[height=6.8cm]{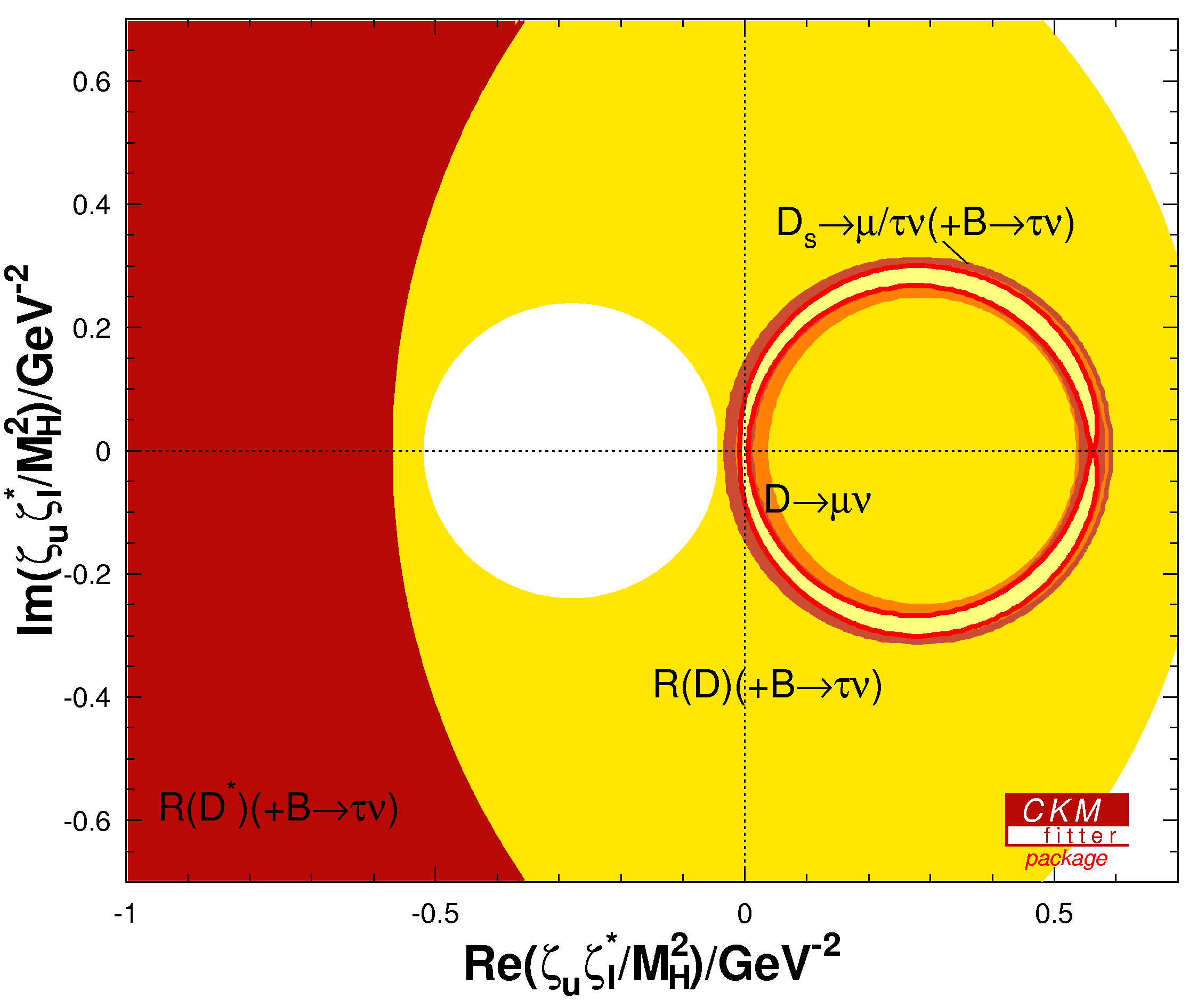}
\caption{\label{zetadlul} \it \small Constraints in the complex $\varsigma_d \varsigma_l^*/M^2_{H^\pm}$~(left) and $\varsigma_u \varsigma_l^*/M^2_{H^\pm}$~(right) planes, in units of $GeV^{-2}$, from the various semileptonic and leptonic decays. Allowed regions are shown at $95\%$~CL for different combinations of the observables.}
\end{figure}

In the second and third plot in this figure we show the corresponding allowed parameter regions in the complex $\delta_{cb}^l$~(left) and $\Delta_{cb}^l$~(right) planes at $95\%$~CL. Here the strong influence of the leptonic decays becomes visible again, excluding most of the parameter space of $R(D)$ in the $\delta_{cb}^l$ plane, and driving the fit far away from the region indicated by the $R(D^*)$ measurement in the $\Delta_{cb}^l$ plane. To examine this effect further, we plot the individual constraints in the $\varsigma_l^*\varsigma_{u,d}/M_{H^\pm}^2$ planes in Fig.~\ref{zetadlul}.
Here it is seen explicitly that the conflict lies mainly between the leptonic charm decays and $R(D^*)$.
It is also noted that, in order to accommodate the data in scenarios~1 and 2, a large value for $|\varsigma_u \varsigma_l^*|/M^2_{H^\pm} \sim O(10^{-1})$~GeV$^{-2}$ is needed. This is not a direct problem in these scenarios, as the different couplings are not related. It would however point to a very strong hierarchy between the charm and the top couplings or a very large value of the leptonic coupling, as the constraints from observables involving loops like $\mathrm{Br}(Z\to \bar{b}b)$ imply $|\varsigma_u|\lesssim1$ for the top coupling \cite{Jung:2010ik}. In the A2HDM, the combined constraints from leptonic $\tau$ decays and these processes require a small value for $|\varsigma_u \varsigma_l^*|/M^2_{H^\pm}$: under the assumptions that $|\varsigma_d|<50$ and the charged-scalar effects dominate the NP contributions to $Z\to b \bar b$, $|\varsigma_u \varsigma_l^*|/M^2_{H^\pm} < 0.005$~GeV$^{-2}$ was obtained in \cite{Jung:2010ik}. Similar bounds also arise from considering the CP-violating parameter $\epsilon_K$ in $K^0$--$\bar K^0$ mixing and the mass difference $\Delta m_{B^0}$ in $B^0$--$\bar B^0$ mixing~\cite{Jung:2010ik}. This is however compatible with our results above, see Fig.~\ref{zetadlul}. As the focus in this article lies on tree-level contributions, and loop induced quantities have a higher UV sensitivity, we refrain from including these constraints explicitly here.

%%%%%%%%%%%%%%%%%%%%%%%%%%%%%%%%%%%%%%%%%%%%%%%%%%%%%%%%%%%%%%%%%%%%%%%%%%
%%%%%%%%%%%%%%%%%%%%%%%%%%%%%%%%%%%%%%%%%%%%%%%%%%%%%%%%%%%%%%%%%%%%%%%%%%

Having in mind the above scenarios, our main concern is whether they can be differentiated with forthcoming data. In addition, the basic assumption of only scalar NP contributions in these decays can be questioned. We identify several combinations of observables which will signal the presence of additional contributions. In the following section we discuss how future measurements of additional observables in $B\to D^{(*)}\tau\nu$ decays, especially their differential distributions, will provide useful information to address these questions.

\section{Observables sensitive to scalar contributions}
\label{sec:observables}

We now proceed to analyze the additional observables in $B\to D^{(*)} \tau \nu_\tau$ decays that provide an enhanced sensitivity to scalar NP contributions. Apart from the differential rates, these are the forward-backward and $\tau$-spin asymmetries in the considered decays, and for $B\to D^*\tau\nu$ additionally the longitudinal polarization fraction of the $D^*$. These observables have been considered in the past by various authors~\cite{Korner:1989qb,Hagiwara:1989gza,Tanaka:1994ay,Chen:2005gr,Chen:2006nua,
Nierste:2008qe,Tanaka:2010se,Fajfer:2012vx,Sakaki:2012ft,Datta:2012qk}, addressing their sensitivity to NP contributions.

Experimentally, while one expects more information on $B\to D^{(*)} \tau \nu_\tau$ decays in the near future from Belle, BaBar, and also LHCb, most of these observables will be accessible only at a Super-Flavour factory~(SFF)~\cite{Aushev:2010bq,O'Leary:2010af}, as their study requires more statistics than the branching ratios and the inclusion of the information from the correlated $B$ meson. The precise sensitivity of future experiments to the different observables has, however, not yet been determined.

For simplicity, we discuss below these additional observables without considering the subsequent decays of the final $\tau$ and $D^*$. While studying the $q^2$ spectra of the observables has the advantage in identifying potential NP contributions and their Dirac structure, this generally requires very high statistics, which might not be available even in the early stages of a SFF. Therefore, we shall analyze both the $q^2$ spectra and the $q^2$-integrated observables. Our predictions for the latter are given in Table~\ref{tab::NOSM}, both within the SM and in the three different scenarios defined before. Note that we do not consider isospin breaking; the observables shown are always isospin-averaged. Note furthermore
that model-independent analyses similar to our scenario 1 have been performed in Refs.~\cite{Fajfer:2012vx,Datta:2012qk,Sakaki:2012ft}.

%%%%%%%%%%%%%%%%%%%%%%%%%%%%%%%%%%%%%%%%%%%%%%%%%%%%%%%%%%%%%%%%%%%%%%%%%%
\begin{table}[t]
\begin{center}
\caption{\label{tab::NOSM} \it \small Predictions for the $q^2$-integrated observables both within the SM and in the different scenarios. The observables have been integrated from $q^2_{\text{min}} = m^2_{\tau}$ to $q^2_{\text{max}}= (m_B - m_{D^{(*)}})^2$. The first error given corresponds to the statistical uncertainty, and the second, when given, to the theoretical one.}
\vspace{0.2cm}
\doublerulesep 0.8pt \tabcolsep 0.04in
\small{
\begin{tabular}{l c c c c}\hline\hline
Observable	        &   SM Prediction                       &   Scenario~1	
                    &   Scenario~2	                        &   Scenario~3 \\ \hline
$ R_L(D^*)$ 	        & $0.115 \pm 0.001 \pm 0.003$           & $0.217\pm0.026$
                    & $0.223^{+0.013}_{-0.026}\pm0.006$	    & $0.104^{+0.006}_{-0.003}\pm0.003$\\

$A_{\lambda}({D})$ 	& $-0.304\pm0.001\pm0.035$				& $-0.55^{+0.10}_{-0.04}$			
                    & $-0.55^{+0.09}_{-0.04}$			    & $-0.55^{+0.09}_{-0.04}\pm0.01$\\
$A_{\lambda}({D^*})$ & $0.502^{+0.005}_{-0.006}\pm 0.017$    & $0.06^{+0.10}_{-0.06}$			
                    & $0.04^{+0.10}_{-0.03}\pm0.01$	    	& $0.57^{+0.04}_{-0.02}\pm0.02$\\
$A_{\theta}({D})$ 	& $0.3602^{+0.0006}_{-0.0007}\pm 0.0022$& $0.03^{+0.01}_{-0.00}\pm0.30$		
                    & $-0.21^{+0.13}_{-0.00}\pm0.06$	    & ${0.36^{+0.01}_{-0.09}}^\dagger$\\
$A_{\theta}({D^*})$	& $-0.066\pm 0.006 \pm 0.009 $          & $-0.136^{+0.012}_{-0.003}\pm0.222$
                    & $0.081^{+0.008}_{-0.059}\pm0.009$	    & $-0.146^{+0.039}_{-0.017}\pm0.021$\\
\hline\hline
\end{tabular}
\footnotesize \parbox[thb]{15.5cm} {\vspace{0.15cm} ${}^\dagger$ Note that the lower tail is rather long, due to a suppressed local maximum.}}
\end{center}
\end{table}
%%%%%%%%%%%%%%%%%%%%%%%%%%%%%%%%%%%%%%%%%%%%%%%%%%%%%%%%%%%%%%%%%%%%%%%%%%

\subsection{The differential decay rates}
\label{subsec:dBr-dRs}

First, we obtain the singly differential rates by summing in Eqs.~\eqref{eq:double-B2D} and \eqref{eq:double-B2Dstar} over the $\tau$ helicities, $\lambda_{\tau}=\pm 1/2$, and performing the integration over $\cos\theta$:
\begin{align} \label{eq:dBrs}
\frac{d\Gamma(\bar B\to D\tau^-\bar\nu_\tau)}{d q^2} &= \frac{G_F^2 |V_{cb}|^2 |\vec{\textbf{p}}| q^2 }{96 \pi^3 m_B^2}\, \left( 1 - \frac{m^2_{\tau}}{q^2} \right)^2\, \left[ |H_{0}|^2\, \left( 1 + \frac{m^2_{\tau}}{2 q^2} \right) + \frac{3 m^2_{\tau}}{2 q^2}\, |H_{t}|^2 \right] \,, \\[0.2cm]
\frac{d\Gamma(\bar B\to D^{*}\tau^-\bar\nu_\tau)}{d q^2} &= \frac{G_F^2 |V_{cb}|^2 |\vec{\textbf{p}}| q^2 }{96 \pi^3 m_B^2}\, \left( 1 - \frac{m^2_{\tau}}{q^2} \right)^2\, \left[ \left( |H_{++}|^2 + |H_{--}|^2 + |H_{00}|^2 \right)\, \left( 1 + \frac{m^2_{\tau}}{2 q^2} \right)\, \right. \nonumber \\
& \left. \qquad + \frac{3 m^2_{\tau}}{2 q^2}\, |H_{0t}|^2 \right] \,.
\end{align}
Again, normalizing to the decays with light leptons reduces the theoretical error:
\begin{equation} \label{eq:dRs}
R_{D^{(*)}}(q^2) = \frac{d\Gamma(\bar B\to D^{(*)} \tau^-\bar\nu_\tau)/d q^2}{d\Gamma(\bar B\to D^{(*)} \ell^-\bar\nu_{\ell})/d q^2} \,.
\end{equation}
Note that in order to obtain the expression for $R_{D^{(*)}}$ from this, numerator and denominator have to be integrated separately.
This should be kept in mind as well for the other quantities.

%%%%%%%%%%%%%%%%%%%%%%%%%%%%%%%%%%%%%%%%%%%%%%%%%%%%%%%%%%%%%%%%%%%%%%%%%%
\begin{figure}[thb]
\centering
\includegraphics[width=7.8cm,height=5.0cm]{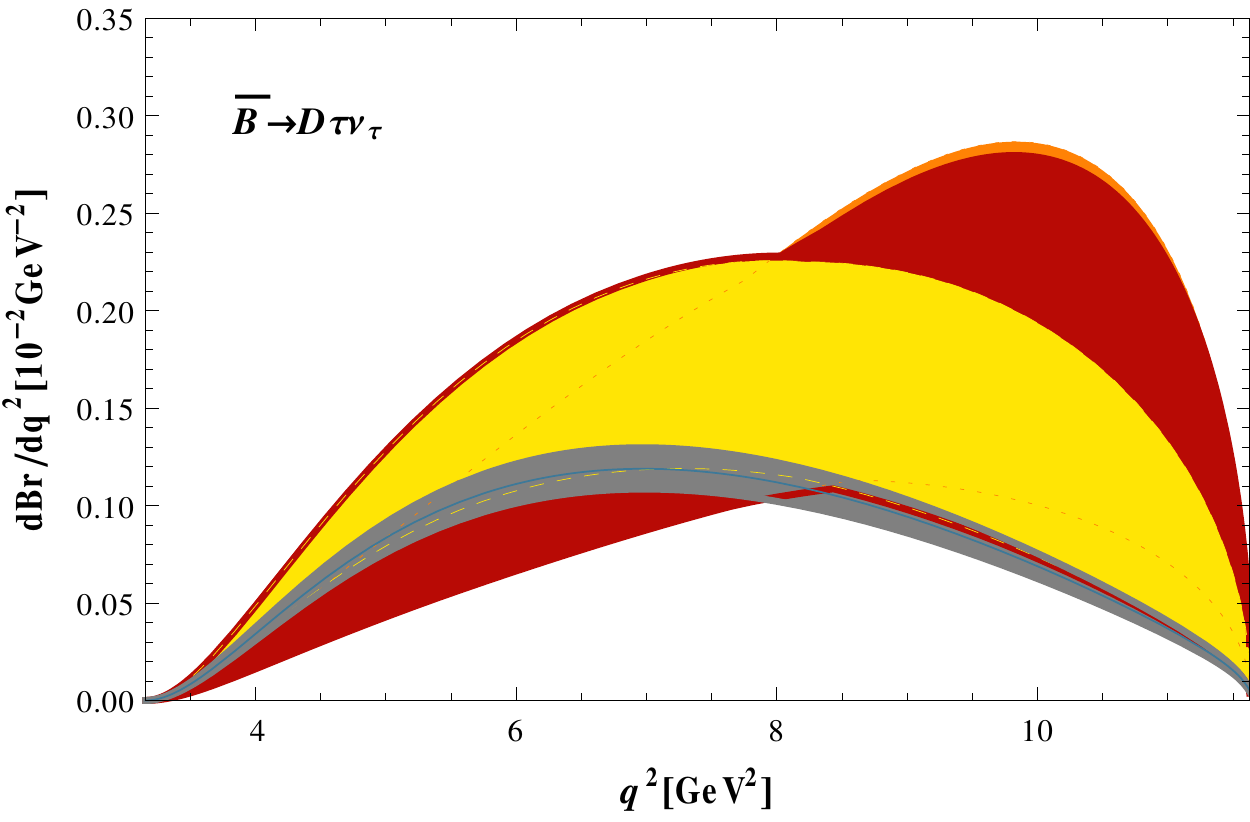}
\hspace{0.5cm}
\includegraphics[width=7.8cm,height=5.0cm]{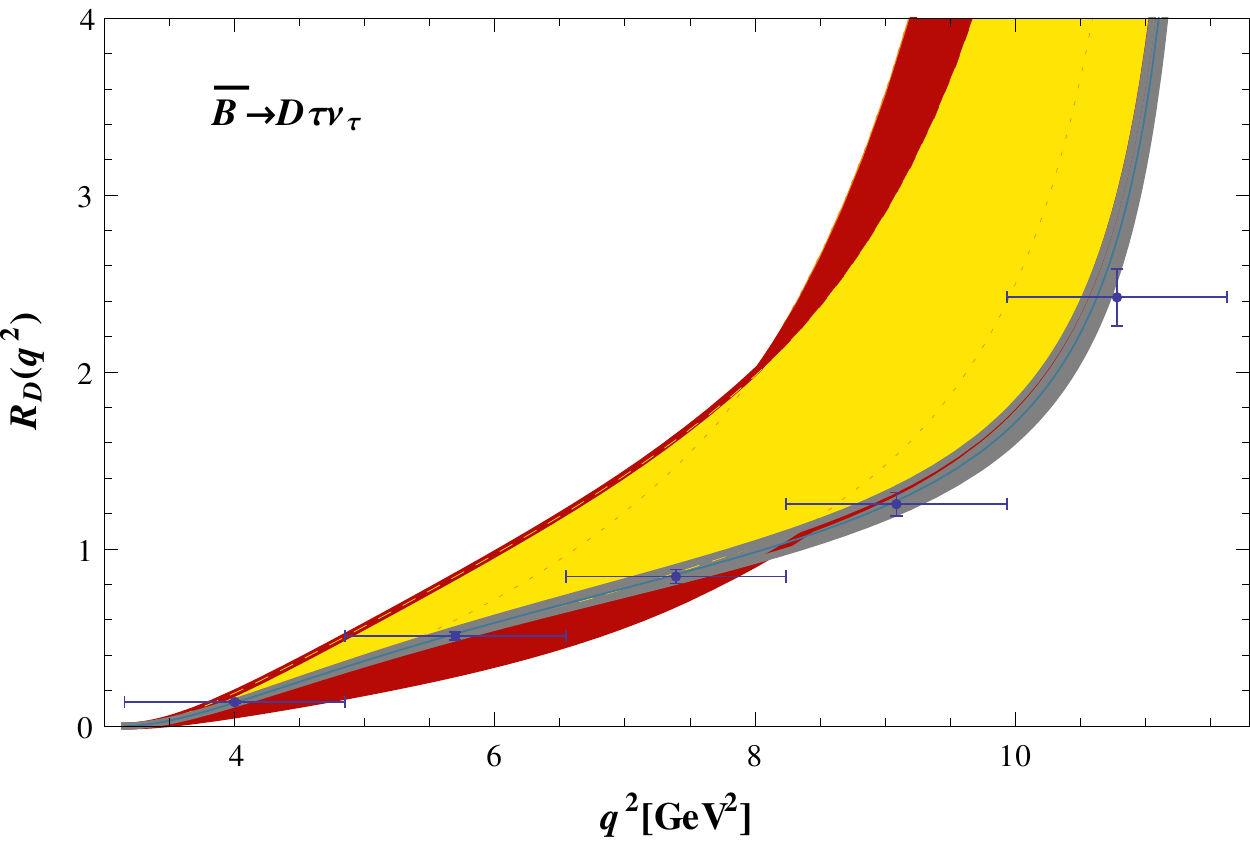}\\[0.2cm]
\includegraphics[width=7.8cm,height=5.0cm]{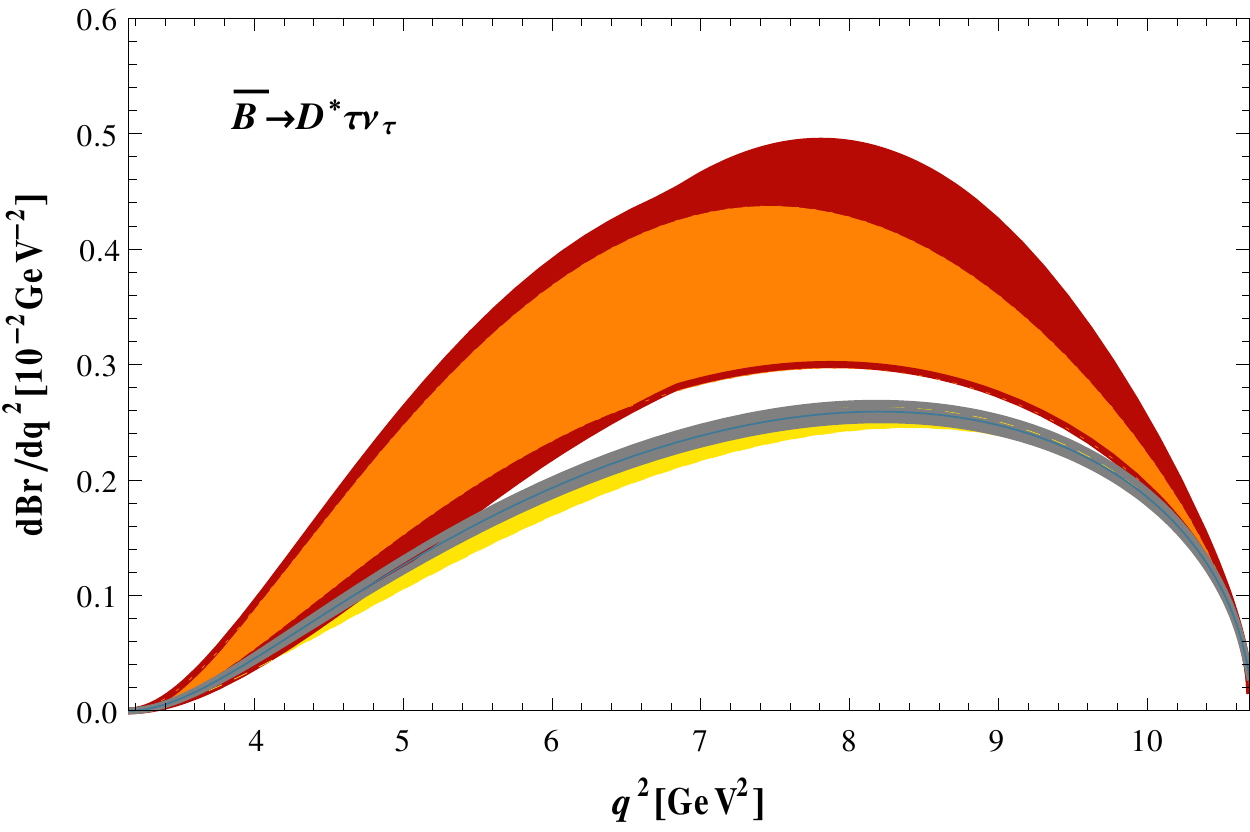}
\hspace{0.5cm}
\includegraphics[width=7.8cm,height=5.0cm]{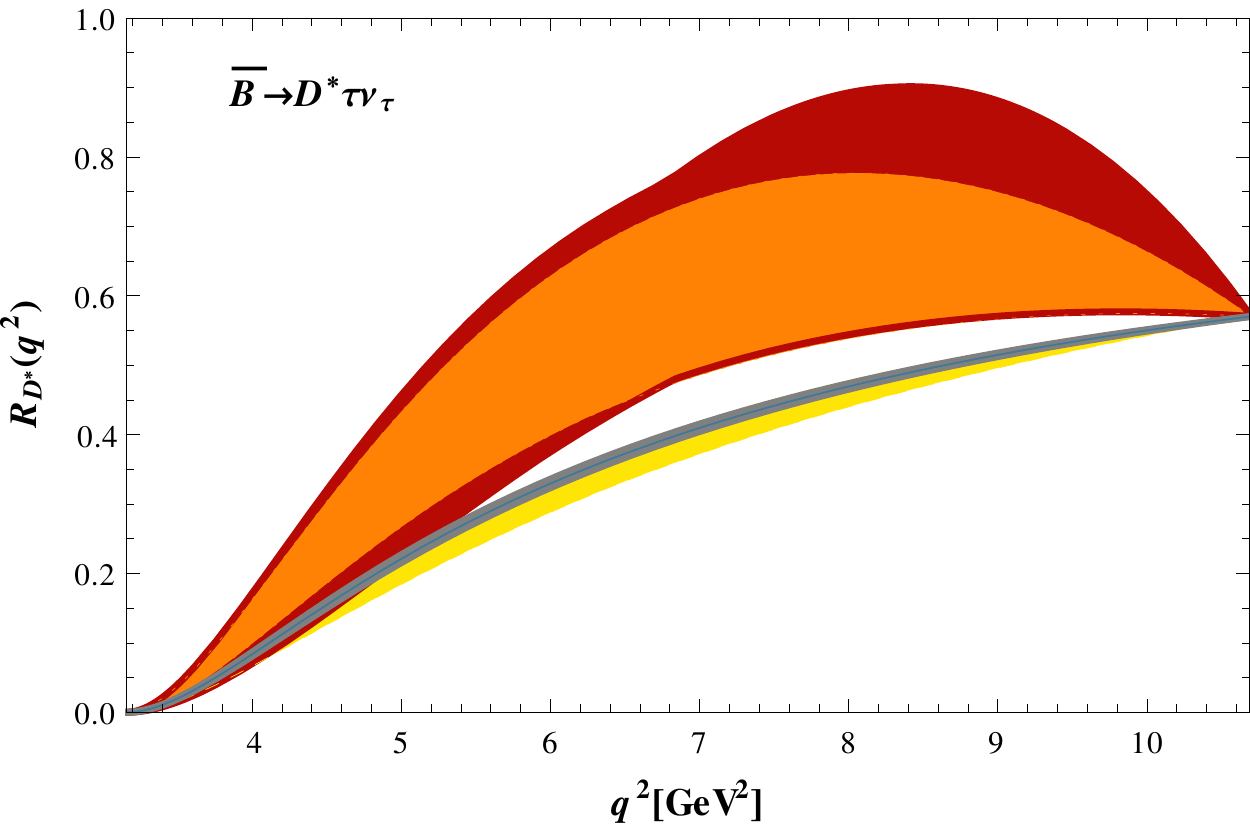}
\caption{\label{fig:dBrs-DRs} \it \small The $q^2$ dependence of the differential branching ratios~(left) and $R_{D^{(*)}}(q^2)$~(right), both within the SM~(grey) as well as in scenario~1~(red), scenario~2~(orange), and scenario~3~(yellow). The binned distribution for $R_{D}(q^2)$ is also shown.}
\end{figure}
%%%%%%%%%%%%%%%%%%%%%%%%%%%%%%%%%%%%%%%%%%%%%%%%%%%%%%%%%%%%%%%%%%%%%%%%%%

Our predictions for these observables, within the SM and in the three different scenarios, are shown in Fig.~\ref{fig:dBrs-DRs}. For $R_D(q^2)$, we show in addition the binned distribution in five equidistant $q^2$ bins, as the ratio does diverge at the endpoint, where however both rates vanish.
From these plots, we make the following observations:

\begin{enumerate}

\item[$\bullet$] As expected, the uncertainty due to the hadronic form factors~(the grey shaded band) in $R_{D^{(*)}}(q^2)$ is significantly reduced compared to that in the differential branching ratio.

\item[$\bullet$] In scenario~2, relatively large deviations from the SM are predicted for almost the full range in $R_{D^*}(q^2)$ and for high $q^2$ in $R_D(q^2)$. The predicted $q^2$ spectra in scenario~3 are close to those of the SM, especially for the $D^{*}$ decay mode. However, there is still room for differences, especially in $R_D(q^2)$, and furthermore in this scenario the distribution in $R_{D^*}(q^2)$ lies preferably \emph{below} the SM one, which should differentiate this scenario clearly from the first two.

\item[$\bullet$] Compared to the SM prediction, the peak of the differential branching ratio in the NP scenarios~(especially in scenario~2) is shifted to higher and lower values in $q^2$ for the $D$ and the $D^{*}$ decay mode, respectively. This is characteristic of scalar NP contributions and should allow for a separation from models with different Dirac structure. The reason for that is the following: while both of them are explicitly proportional to $q^2$~(see Eqs.~\eqref{eq:BD-helicities} and \eqref{eq:BDstar-helicities}), the charged-scalar contribution to $R_{D^{*}}(q^2)$ is in addition proportional to the $D^*$ momentum $|\vec{\textbf{p}}|$, which vanishes at the endpoint $q^2_{\text{max}}=(m_B-m_{D^{*}})^2$, rendering its relative contribution maximal for intermediate values of $q^2$, while the one to $R_{D}(q^2)$ continuously increases with $q^2$. The relative suppression of the terms proportional to the $\tau$ mass by the $D^*$ momentum furthermore renders $R_{D^{*}}(q^2)$ finite everywhere, while $R_{D}(q^2)$ diverges at the endpoint.\footnote{In fact, this behavior is an artifact of setting $m_{\ell}\equiv0$. The actual value is $\sim m_\tau^2/m_\ell^2$.} However, as both the rates for the $\tau$ and the light lepton modes vanish there, this does not influence the experimental extraction: when calculating the contributions to different bins, all integrals remain finite. These characteristic features are illustrated by the right two plots.

\end{enumerate}

\begin{figure}[thb]
\centering
\includegraphics[width=7.8cm,height=5.0cm]{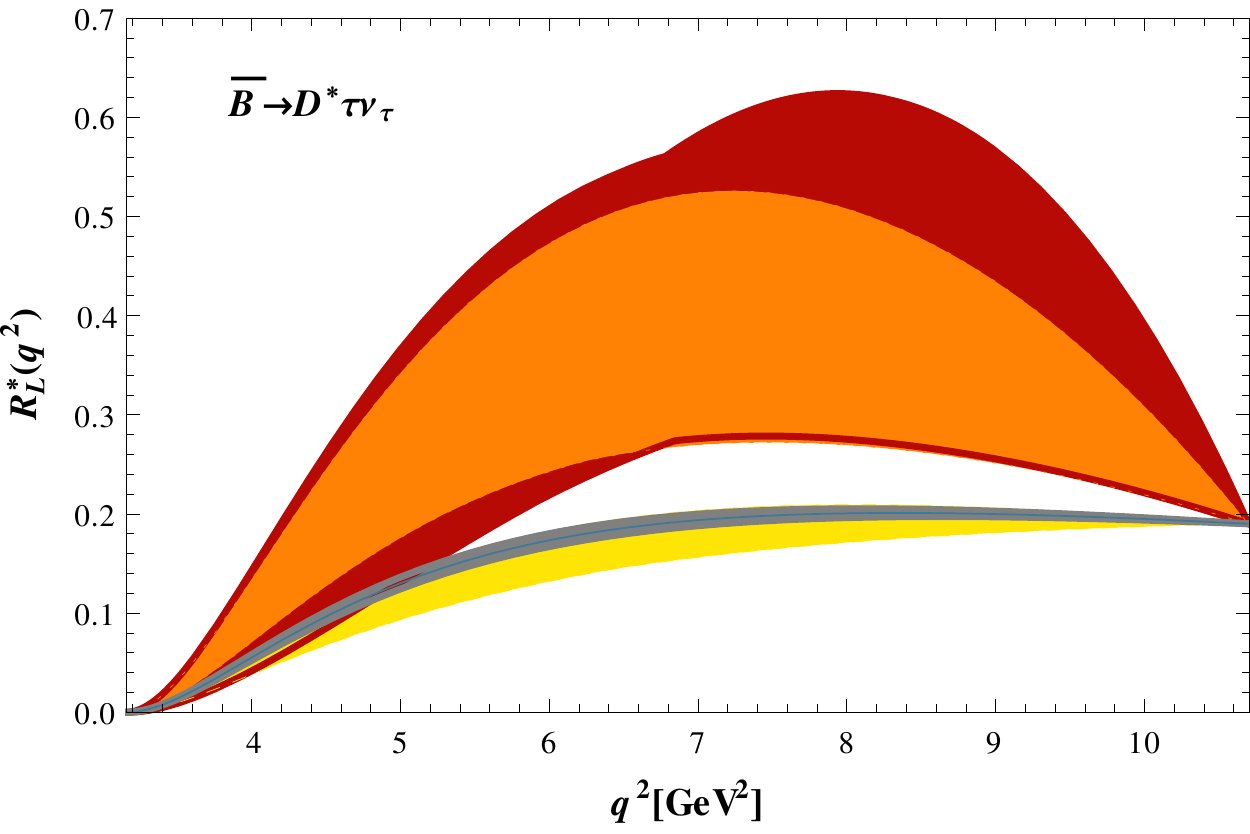}
\caption{\label{fig:DRL} \it \small Predictions for $R_L^*(q^2)$ both within the SM and in the three different scenarios. The other captions are the same as in Fig.~\ref{fig:dBrs-DRs}.}
\end{figure}

As the charged scalar only contributes to the helicity amplitude $H_{0t}$ in $B\to D^{*} \tau \nu_\tau$ decays, an increased sensitivity is expected by studying the case with a longitudinally polarized $D^*$ meson in the final state, where the transverse helicity amplitudes are no longer relevant. For this purpose, we define a singly differential longitudinal decay rate~\cite{Fajfer:2012vx}:
\begin{equation}
\frac{d \Gamma^{L}_{\tau}}{d q^2} = \frac{G_F^2 |V_{cb}| |\vec{\textbf{p}}| q^2}{96 \pi^3 m_B^2}\, \left( 1 -  \frac{m_{\tau}^2}{q^2} \right)^2\, \left[ |H_{00}|^2\, \left( 1 + \frac{m_{\tau}^2}{2 q^2} \right) + \frac{3 m_{\tau}^2}{2 q^2}\, |H_{0t}|^2 \right] \,.
\end{equation}
In analogy to $R_{D^{*}}(q^2)$, it is again advantageous to consider the ratio with the $\tau$ mode normalized to the light lepton mode:
\begin{equation}
R_L^*(q^2) =  \frac{d \Gamma^{L}_{\tau}/d q^2}{d \Gamma^{L}_{\ell}/d q^2} \,.
\end{equation}
It is important, however, to note that within our NP framework this is not an independent observable. As long as we consider only additional scalar operators, the difference
\begin{equation}\label{eq::X1}
X_1(q^2)\equiv R_{D^*}(q^2)-R_L^*(q^2)
\end{equation}
is independent of NP effects. A measurement of this observable serves, therefore, as a cross-check for the effect in $R_{D^*}$ and gives us information on whether scalar NP operators are sufficient to describe the data. This observation is reflected in Table~\ref{tab::NOSM} and in Fig.~\ref{fig:DRL}, where we show the predictions for $R_L(D^*)$ and $R_L^*(q^2)$ both within the SM and in the three different scenarios; the results are analogous to the ones for $R({D^*})$ and $R_{D^*}(q^2)$ discussed above, but clearly exhibit an increased sensitivity to the scalar NP effect.

%%%%%%%%%%%%%%%%%%%%%%%%%%%%%%%%%%%%%%%%%%%%%%%%%%%%%%%%%%%%%%%%%%%%%%%%%%
%%%%%%%%%%%%%%%%%%%%%%%%%%%%%%%%%%%%%%%%%%%%%%%%%%%%%%%%%%%%%%%%%%%%%%%%%%

\subsection{The $\tau$ spin asymmetry}
\label{subsec:Alambda}

\begin{figure}[thb]
\centering
\includegraphics[width=7.8cm,height=5.0cm]{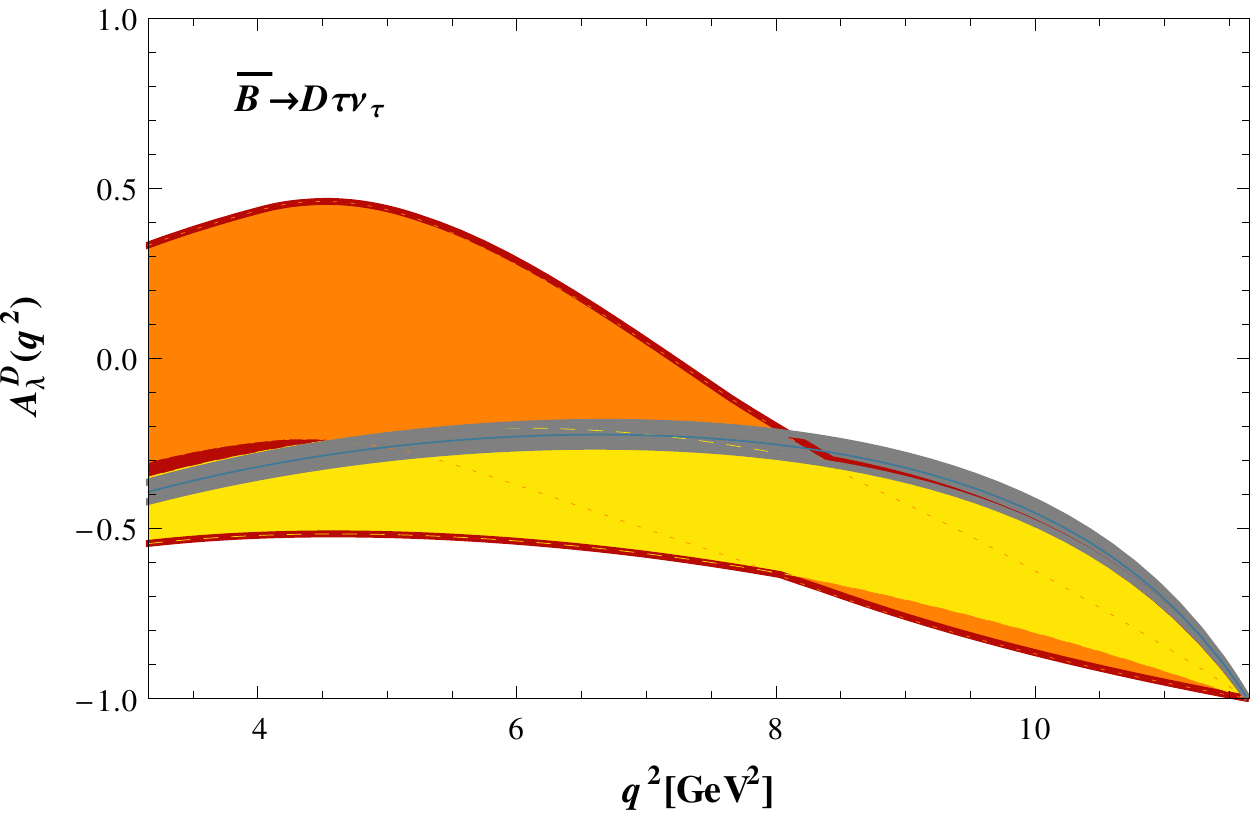}
\hspace{0.5cm}
\includegraphics[width=7.8cm,height=5.0cm]{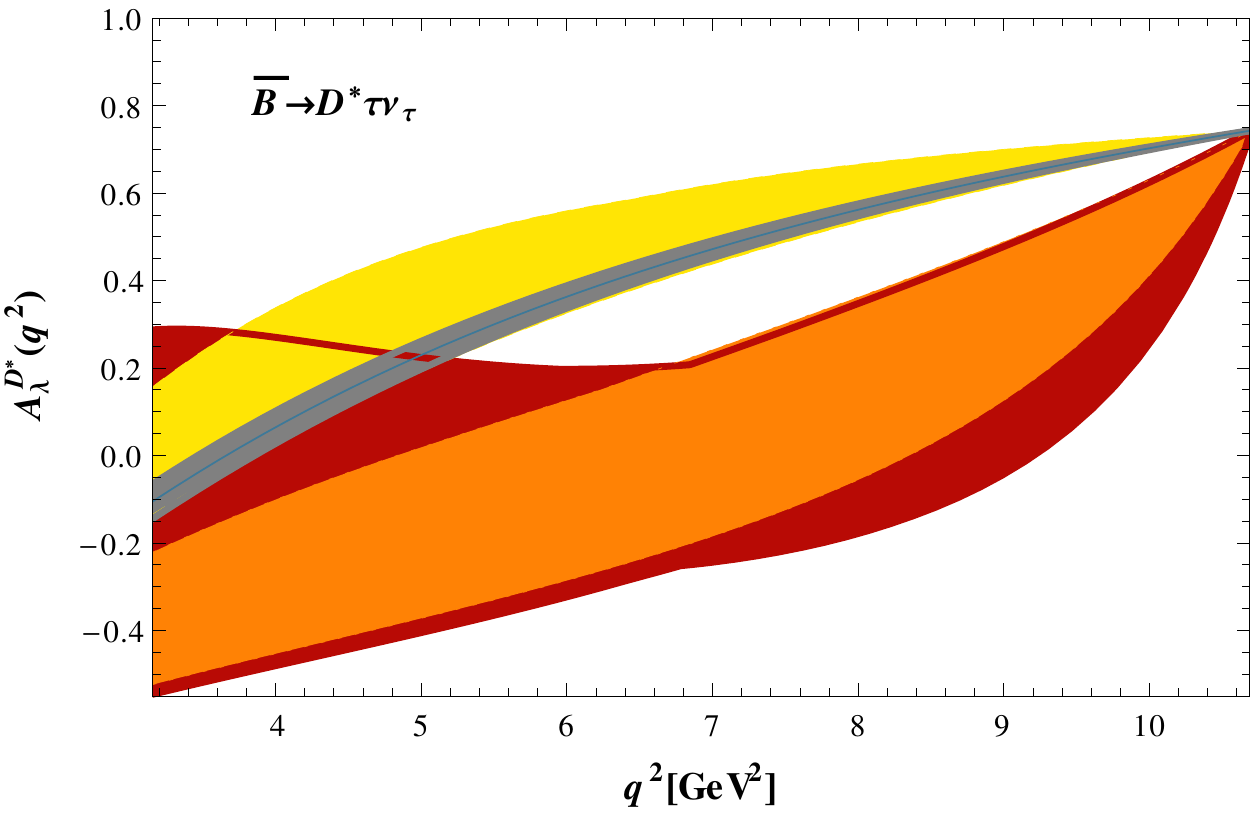}
\caption{\label{fig:Alambda} \it \small The $q^2$ dependence of the $\tau$ spin asymmetries $A^{D}_{\lambda}(q^2)$~(left) and $A^{D^*}_{\lambda}(q^2)$~(right). The other captions are the same as in Fig.~\ref{fig:dBrs-DRs}.}
\end{figure}

\begin{figure}[tbh]
\centering
\includegraphics[width=7.8cm,height=5.0cm]{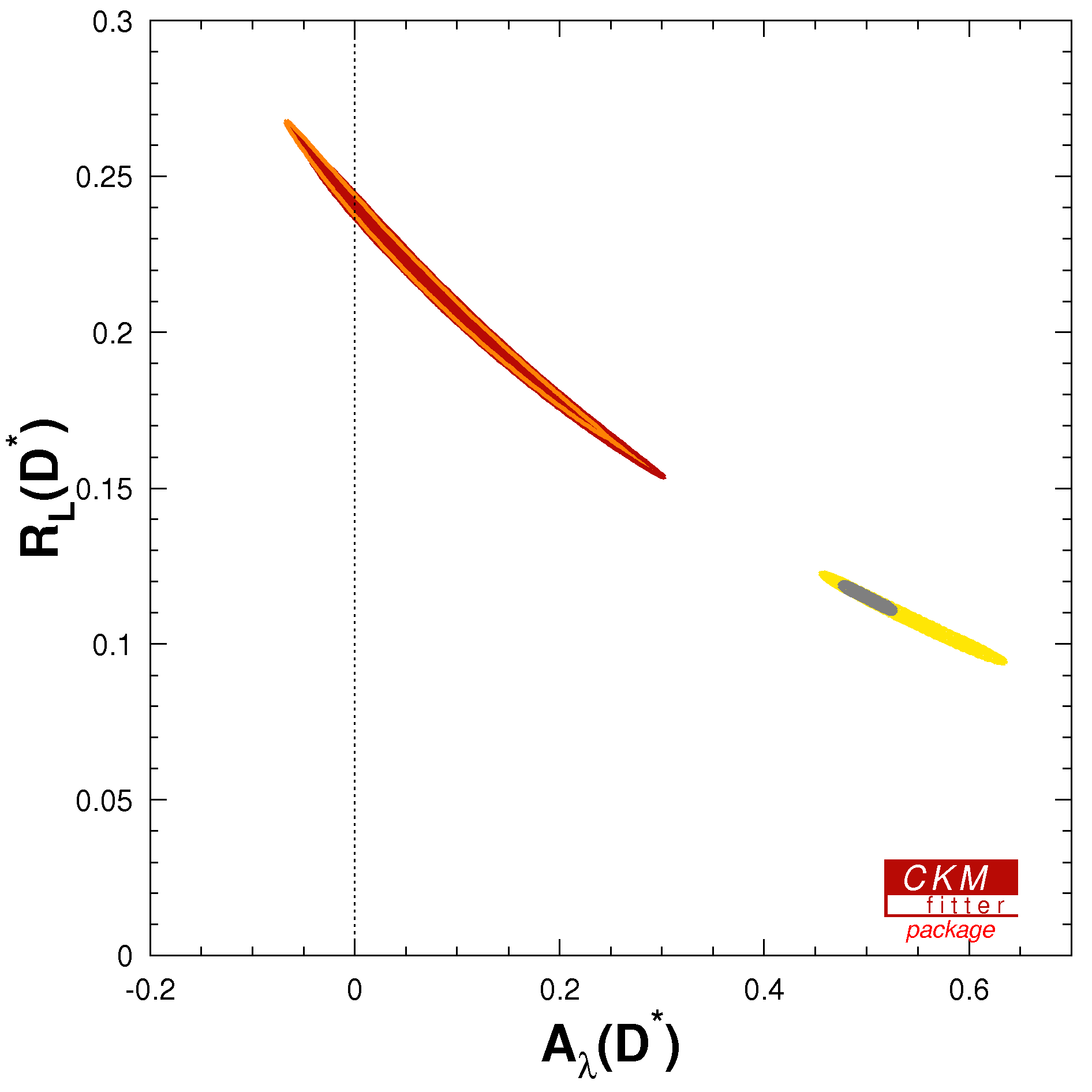}
\caption{\label{fig::AlambdaRDstarL} \it \small Predictions for $A_{\lambda}(D^*)$ vs. $R_L(D^*)$ both within the SM~(grey) and in the three scenarios~((1)-red, (2)-orange, (3)-yellow), from a global fit including the appropriate observables.}
\centering
\end{figure}

Information on the $\tau$ spin in semileptonic $B$-meson decays can be inferred from its distinctive decay patterns~\cite{Tanaka:1994ay,Tanaka:2010se,Fajfer:2012vx,Datta:2012qk}. Therefore, we consider here the $\tau$ spin asymmetry defined in the $\tau$-$\bar\nu_{\tau}$ center-of-mass frame:
\begin{equation} \label{eq:ATAUD}
A^{D^{(*)}}_{\lambda}(q^2) = \frac{ d\Gamma^{D^{(*)}}[\lambda_{\tau} =  -1/2]/ d q^2  - d\Gamma^{D^{(*)}}[\lambda_{\tau} = + 1/2] /d q^2 }{ d\Gamma^{D^{(*)}}[\lambda_{\tau} =  -1/2]/ d q^2  + d\Gamma^{D^{(*)}}[\lambda_{\tau} = + 1/2] /d q^2}\,,
\end{equation}
where the polarized differential decay rates are obtained after integration over $\cos\theta$ of the doubly-differential ones given by Eqs.~\eqref{eq:double-B2D} and \eqref{eq:double-B2Dstar}. Using the formulae presented in appendices~\ref{subsec:BD} and \ref{subsec:BDstar}, we obtain explicitly
\begin{align}  \label{eq:ATAUD2}
A^{D  }_{\lambda}(q^2) &= \frac{ |H_0|^2\, ( 1- \frac{m^2_{\tau}}{2 q^2} ) - \frac{3 m_{\tau}^2}{2 q^2}\, |H_{t}|^2}{ |H_0|^2\, (1 + \frac{m^2_{\tau}}{2 q^2} ) + \frac{3 m_{\tau}^2 }{2 q^2}\, |H_{t}|^2 } \,, \nonumber \\[0.2cm]
A^{D^*}_{\lambda}(q^2) &= \frac{ ( |H_{00}|^2 + |H_{++}|^2 + |H_{--}|^2 )\, ( 1 - \frac{m_{\tau}^2}{2 q^2} ) - \frac{3 m_{\tau}^2}{2 q^2}\, |H_{0t}|^2 }{ ( |H_{00}|^2   + |H_{++}|^2 + |H_{--}|^2 )\, ( 1 + \frac{m_{\tau}^2}{2 q^2} ) + \frac{3 m_{\tau}^2}{2 q^2}\, |H_{0t}|^2 } \,.
\end{align}
Again, these two observables have the same dependence on scalar NP contributions as the differential rates; this observation follows from the combinations
\begin{equation}\label{eq::X12}
X_2^D(q^2)\equiv R_D(q^2)\,(A^{D}_{\lambda}(q^2)+1)\quad\mbox{and}\quad X_2^{D^*}(q^2)\equiv R_{D^{*}}(q^2)\,(A^{D^{*}}_{\lambda}(q^2)+1)
\end{equation}
being independent of $\delta_{cb}^{\tau}$ and $\Delta_{cb}^{\tau}$, respectively. However, because of the different normalization and systematics in this case, a future measurement would give important information on the size and nature of NP in $B\to D^{(*)}\tau\nu$  decays: like for $X_1(q^2)$, any deviation from the SM value of these combinations would indicate non-scalar NP. Our predictions for the two asymmetries when integrated over $q^2$ are given again in Table~\ref{tab::NOSM}; the predicted ranges for the differential distributions are shown in Fig.~\ref{fig:Alambda}. The correlation between $A_{\lambda}(D^*)$ and $R_{L}(D^*)$, following from Eqs.~\eqref{eq::X1} and \eqref{eq::X12}, is furthermore illustrated in Fig.~\ref{fig::AlambdaRDstarL}, where we show the predicted values for the two observables for the SM and the three scenarios, yielding in every case a very small band, corresponding to the hadronic uncertainties.

%%%%%%%%%%%%%%%%%%%%%%%%%%%%%%%%%%%%%%%%%%%%%%%%%%%%%%%%%%%%%%%%%%%%%%%%%%
%%%%%%%%%%%%%%%%%%%%%%%%%%%%%%%%%%%%%%%%%%%%%%%%%%%%%%%%%%%%%%%%%%%%%%%%%%

%%%%%%%%%%%%%%%%%%%%%%%%%%%%%%%%%%%%%%%%%%%%%%%%%%%%%%%%%%%%%%%%%%%%%%%%%%
%%%%%%%%%%%%%%%%%%%%%%%%%%%%%%%%%%%%%%%%%%%%%%%%%%%%%%%%%%%%%%%%%%%%%%%%%%

The integrated asymmetries span a rather large range. For $A_\lambda(D)$, due to the common input, a differentiation between the different scenarios seems very difficult, while for $A_\lambda(D^*)$ at least the  separation of the SM and scenario~3 on the one hand and scenarios~1 and 2 on the other hand is very clear. Importantly, the predictions for $A_\lambda(D)$ in all scenarios are clearly negative, providing another option to potentially exclude the SM and only scalar NP at the same time. Also largely negative values for $A_\lambda(D^*)$ are excluded in all scenarios.

The separation of the different models improves, once differential distributions are considered: from Fig.~\ref{fig:Alambda} we observe very distinct patterns for the different scenarios, especially for scenario~2. This scenario will be clearly distinguishable from the SM and scenario~3, once the necessary experimental precision is reached. Scenario~1, on the other hand, has again possibly very large effects, but might be close to any of the other scenarios, including the SM. A characteristic feature of this observable is the zero-crossing point: for $A_\lambda^D(q^2)$, it is absent for the SM and scenario~3, while it appears likely for scenario~2 to have one, and also scenario~1 has that option. Within the SM, the observable $A_{\lambda}^{D^*}(q^2)$ crosses the zero at $q^2 = 3.66 \pm 0.04~{\rm GeV}^2$; compared to the SM case, the zero-crossing point occurs at significantly higher values of $q^2$ for scenario~2, while most likely at lower values of $q^2$ for scenario~3. This indicates that measuring the zero-crossing point of the $\tau$ spin asymmetries can be a useful probe of the flavour structure of the charged scalar interaction.

\subsection{The forward-backward asymmetries}
\label{subsec:Atheta}

\begin{figure}[thb]
\centering
\includegraphics[width=7.8cm,height=5.0cm]{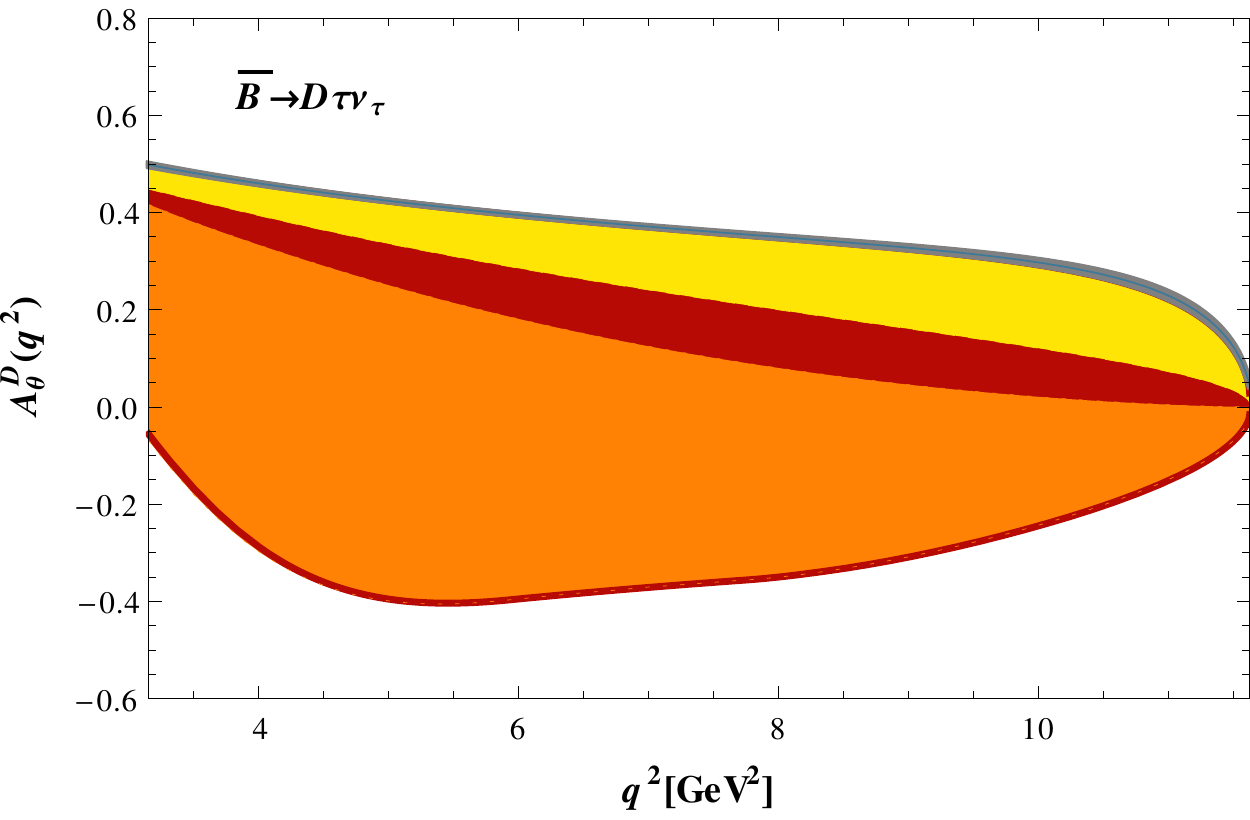}
\hspace{0.5cm}
\includegraphics[width=7.8cm,height=5.0cm]{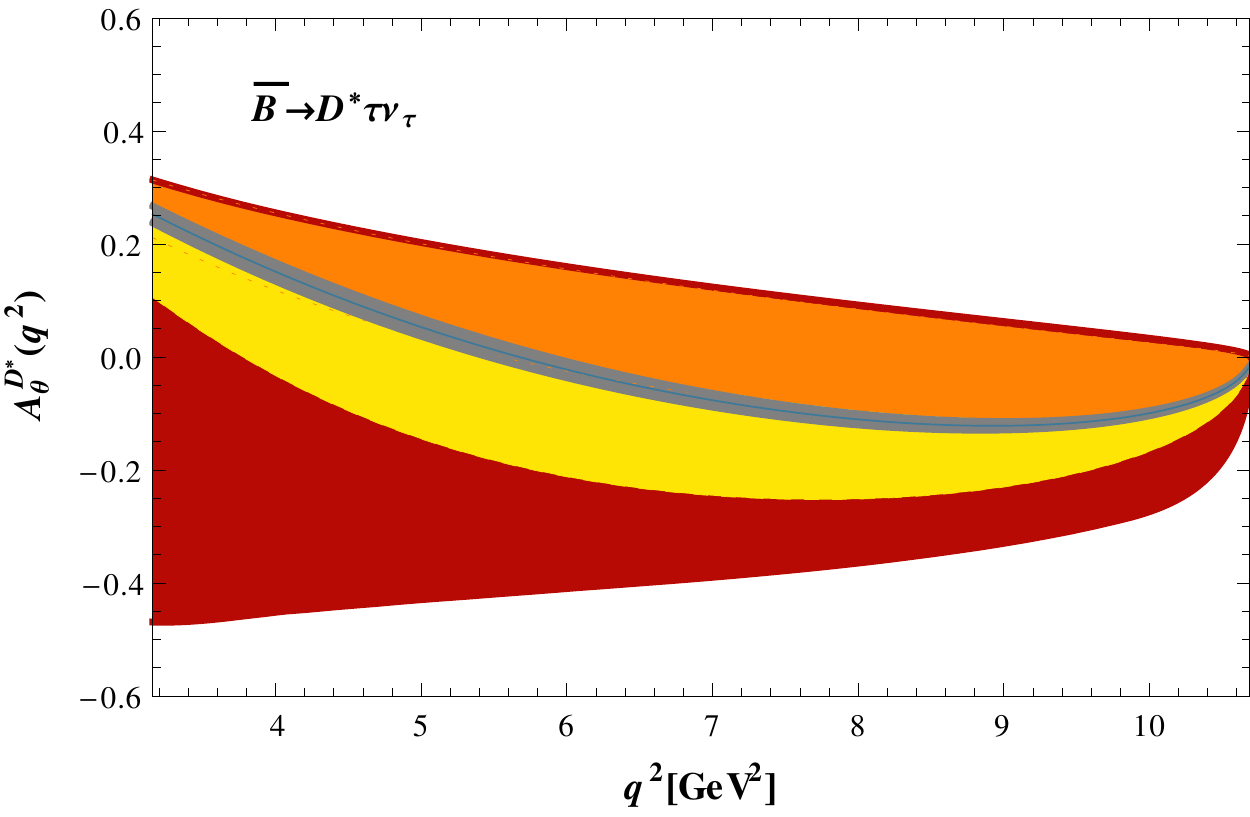}
\caption{\label{fig:Atheta} \it \small The $q^2$ dependence of the forward-backward asymmetries $A^{D}_{\theta}(q^2)$~(left) and $A^{D^*}_{\theta}(q^2)$~(right). The other captions are the same as in Fig.~\ref{fig:dBrs-DRs}.}
\end{figure}

\begin{figure}[tbh]
\centering
\includegraphics[height=5.7cm]{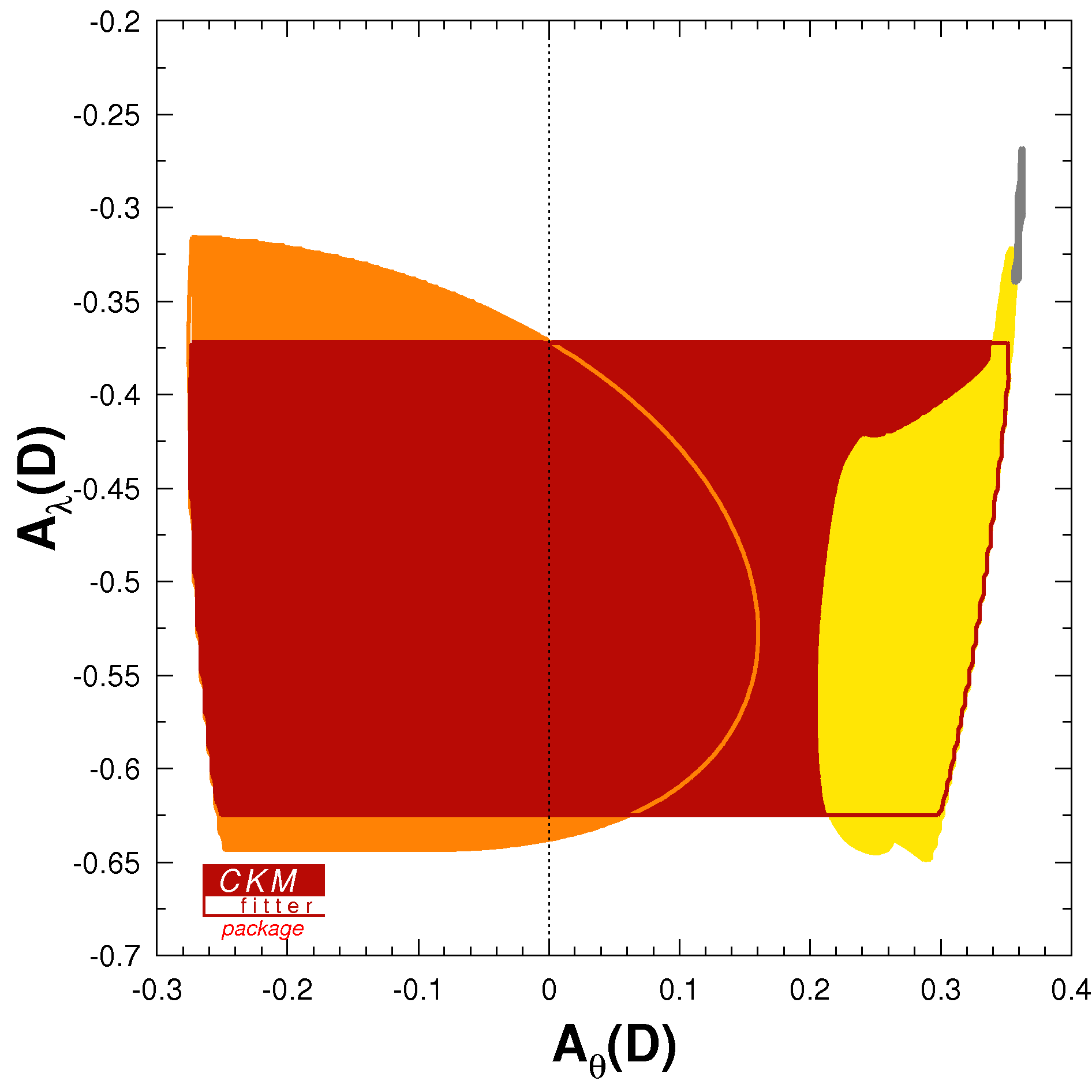}
% \hspace{0.5cm}
\includegraphics[height=5.7cm]{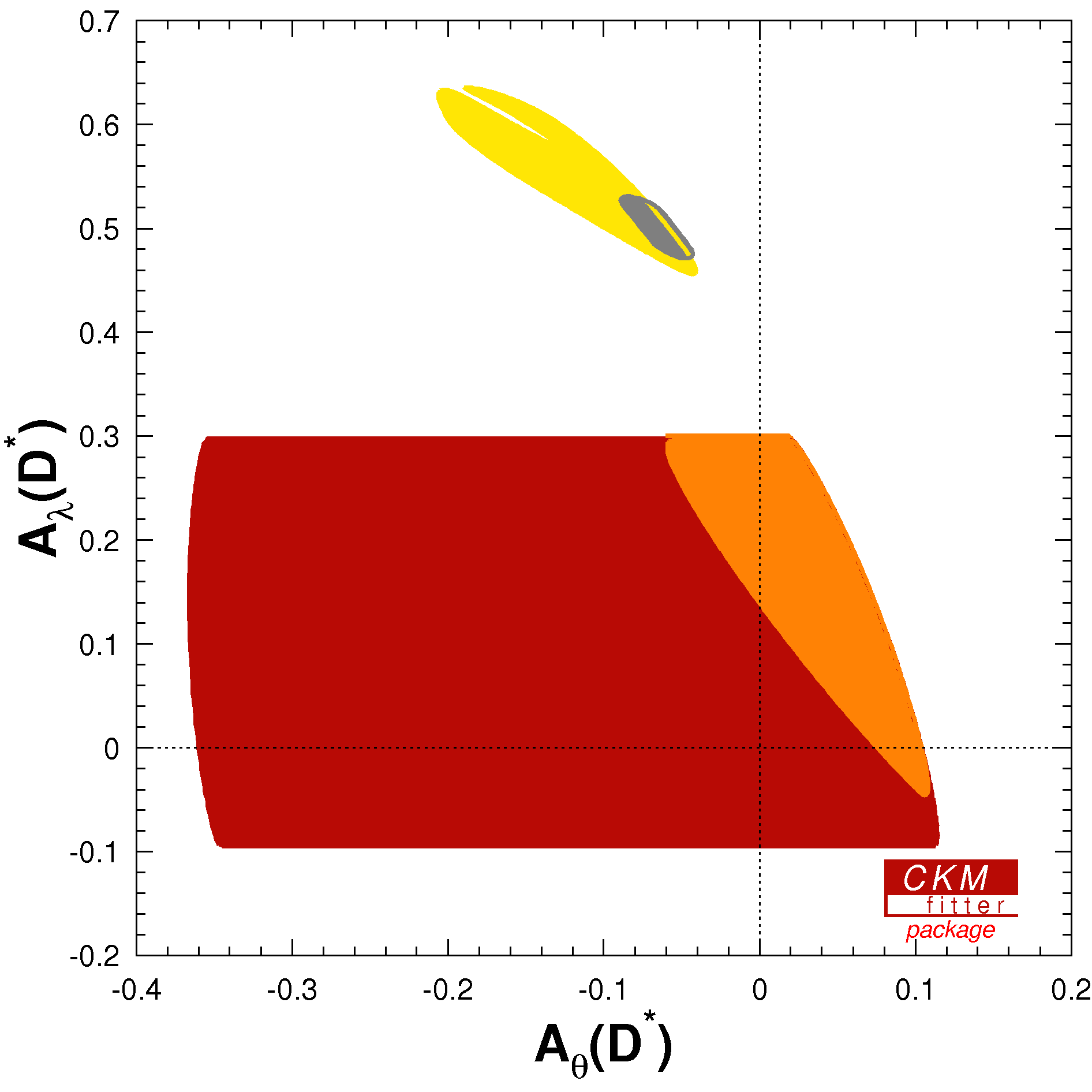}
\includegraphics[height=5.7cm]{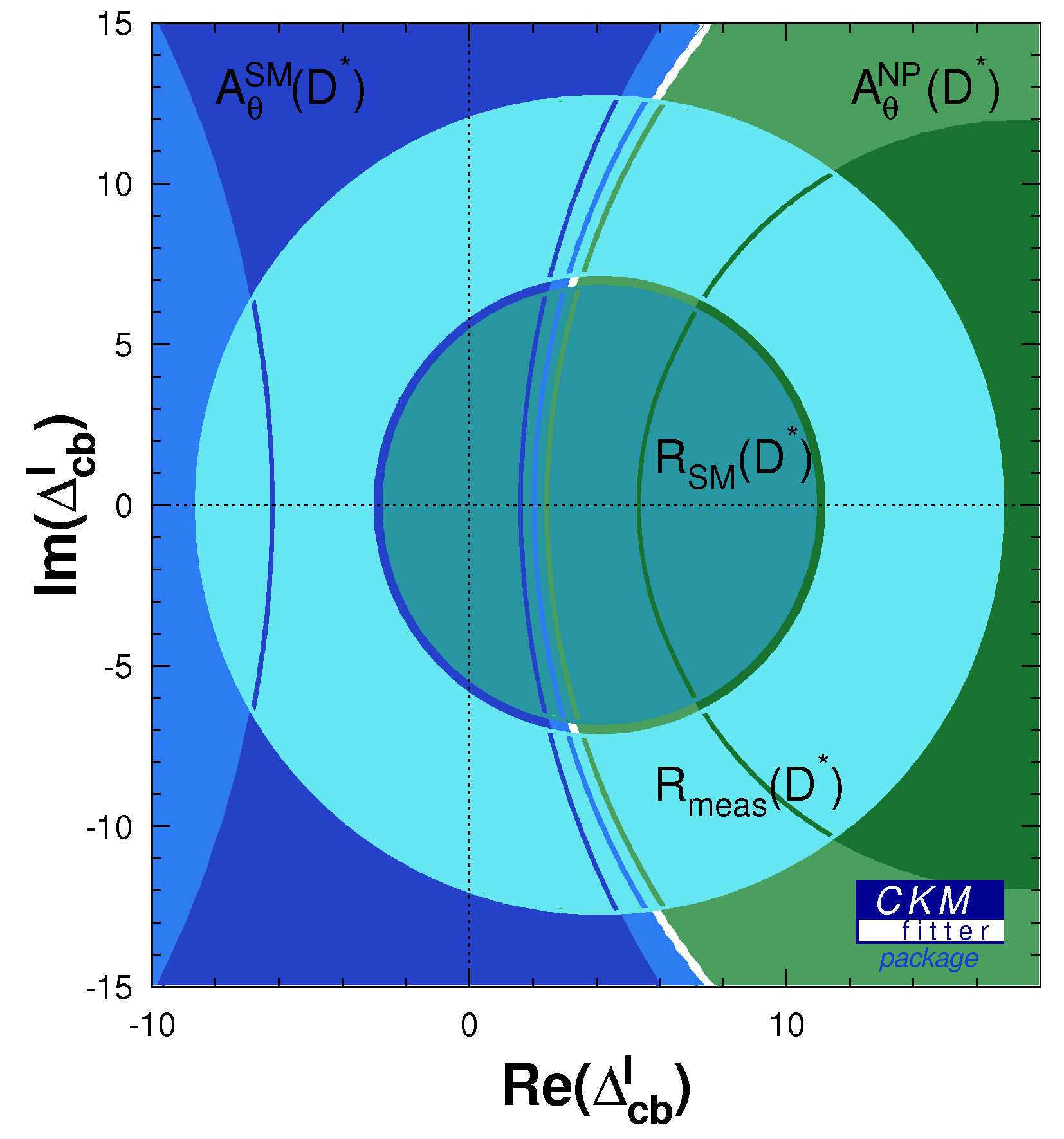}
\caption{\label{fig::AFBAlambda} \it \small Prediction for $A_{\theta}(D^{(*)})$ vs. $A_{\lambda}(D^{(*)})$ for the SM (grey), and the three scenarios ((1)-red, (2)-orange, (3)-yellow), from a global fit including all appropriate observables.     The right plot shows the possible impact of future measurements on the complex $\Delta_{cb}^\tau$ plane (see text).}
\centering
\end{figure}

Finally, we discuss the forward-backward asymmetries defined as the relative difference between the partial decay rates where the angle $\theta$ between the $D^{(*)}$ and $\tau$ three-momenta in the $\tau$-$\bar\nu_{\tau}$ center-of-mass frame is greater or smaller than $\pi/2$:
\begin{align} \label{eq:AFBDs}
A^{D^{(*)}}_{\theta}(q^2) &= \frac{ \int_{-1}^{0} d \cos\theta\, ( d^2 \Gamma^{D^{(*)}}_{\tau}/dq^2 d \cos\theta )  -\int_{0}^{1} d \cos\theta\, ( d^2 \Gamma^{D^{(*)}}_{\tau}/dq^2 d \cos\theta ) }{ d \Gamma^{D^{(*)}}_{\tau}/dq^2 } \,.
\end{align}
Using Eqs.~\eqref{eq:double-B2D} and \eqref{eq:double-B2Dstar}, we arrive at the following explicit expressions:
\begin{align} \label{eq:AFBD}
A^{D  }_{\theta}(q^2) &=\frac{3 m_{\tau}^2}{2 q^2}\, \frac{ \mathrm{Re}( H_0 H_{t}^* ) }{ |H_0|^2\, ( 1 + \frac{m_{\tau}^2}{2 q^2} ) + \frac{3 m_{\tau}^2}{2 q^2}\, |H_{t}|^2} \,, \nonumber \\[0.2cm]
A^{D^*}_{\theta}(q^2) &= \frac{3}{4}\, \frac{ | H_{++}|^2 - |H_{--}|^2 + 2 \frac{m_{\tau}^2}{q^2}\, \mathrm{Re}( H_{00} H_{0t}^* ) }{ (|H_{++}|^2 + |H_{--}|^2 + |H_{00}|^2)\, ( 1+ \frac{m_{\tau}^2}{2 q^2}) + \frac{3 m_{\tau}^2}{2 q^2}\, |H_{0t}|^2 } \,.
\end{align}
In terms of a model-independent determination of NP parameters, {\it i.e.} scenario~1, this is the key observable to determine $\Delta_{cb}^\tau$ and $\delta_{cb}^\tau$. The reason, as mentioned before, is that the observables $R_{L}^*(q^2)$, $R_{D^{(*)}}(q^2)$ and $A_\lambda^{D^{(*)}}(q^2)$ do not give independent information, see Eqs.~\eqref{eq::X1} and \eqref{eq::X12}. The forward-backward asymmetry $A^{D^{(*)}}_{\theta}(q^2)$ is therefore the \emph{only} independent constraint in the complex $\delta_{cb}^{\tau}$~($\Delta_{cb}^{\tau}$) plane. Our predictions for this observable are given in Table~\ref{tab::NOSM} and shown in Fig.~\ref{fig:Atheta}. Furthermore, its correlation with the $\tau$ spin asymmetry for the two modes is shown in the first two  panels in Fig.~\ref{fig::AFBAlambda}. It is clearly seen that the correlation is much weaker in this case, especially in scenario~1, where the only influence stems from the restriction on $|\delta_{cb}^\tau|$ and $|\Delta_{cb}^\tau|$. However, the pattern of a more SM-like A2HDM prediction and strongly shifted predictions from scenarios~1 and 2 is repeated. Regarding the differential distributions, within the SM, the observable $A_{\theta}^D(q^2)$ does not cross zero, while this becomes possible for scenarios~1 and 2. $A_{\theta}^{D^*}(q^2)$ has a zero-crossing point at $q^2 = 5.67 \pm 0.02~{\rm GeV}^2$ in the SM, for which again large shifts are possible with NP, and it might even vanish in scenario~3. Large deviations from the SM expectations, especially in scenario~2, are therefore still possible for this observable.

%%%%%%%%%%%%%%%%%%%%%%%%%%%%%%%%%%%%%%%%%%%%%%%%%%%%%%%%%%%%%%%%%%%%%%%%%%
%%%%%%%%%%%%%%%%%%%%%%%%%%%%%%%%%%%%%%%%%%%%%%%%%%%%%%%%%%%%%%%%%%%%%%%%%%

In order to illustrate the impact of a possible future measurement of this observable, we exemplarily show in the right panel in Fig.~\ref{fig::AFBAlambda} the resulting constraint in the $\Delta_{cb}^\tau$ plane, together with the one from $R(D^*)$ as measured at the moment. The $A_\theta(D^*)$ constraints drawn in lighter colours correspond to an uncertainty of $10\%$. For the darker constraints, an improvement by a factor of 2 has been assumed compared to the lighter ones. Furthermore, an index `SM' indicates the measurement chosen to be compatible with the SM, while the index `NP' corresponds to measurements excluding the SM, but compatible with scenario~1. As can be seen, such a measurement would allow to exclude a large part of the parameter space in the model-independent scenario~1, as well as constrain the other scenarios further. Furthermore, as mentioned before, the two constraints could also miss each other in that plane, indicating NP with a different Dirac structure. This possibility exists of course also for the other observables discussed above.

Additional information on $\Delta_{cb}^\tau$ could obviously be obtained from a measurement of the $B_c^-\to\tau^-\bar\nu_\tau$ rate. With the NP influence being determined by $\Delta_{cb}^\tau$, this rate is clearly predicted to be different from the SM in scenarios~1 and 2, while close to the SM in scenario~3. However, this mode is extremely hard to be measured experimentally.

%%%%%%%%%%%%%%%%%%%%%%%%%%%%%%%%%%%%%%%%%%%%%%%%%%%%%%%%%%%%%%%%%%%%%%%%%%
%%%%%%%%%%%%%%%%%%%%%%%%%%%%%%%%%%%%%%%%%%%%%%%%%%%%%%%%%%%%%%%%%%%%%%%%%%

\section{Summary}
\label{sec:conclusion}
In this paper, motivated by the recent experimental evidence for an excess of $\tau$-lepton production in exclusive semileptonic $B$-meson decays, we have performed a detailed phenomenological analysis of $b\to q\,\tau^-\bar\nu_\tau$~($q=u,c$) transitions within a framework with additional scalar operators, assumed to be generated by the exchange of a charged scalar in the context of 2HDMs.

While the usual Type-II scenario cannot accommodate the recent BaBar data on $\bar B \to D^{(*)} \tau^- \bar\nu_\tau$ decays, this is possible within more general models, in which the charged-scalar couplings to up-type quarks are not as suppressed. An explicit example is given by the A2HDM, in which
the $\bar B \to D^{(*)}\tau^- \nu_\tau$ as well as the $B^-\to\tau^-\bar\nu_\tau$ data can be fitted. However,
the resulting parameter ranges are in conflict with the constraints from leptonic charm decays, which could indicate a departure from the family universality of the Yukawa couplings
$\varsigma_f$ ($f=u,d,l$).

These observations led us to define three scenarios for scalar NP, in which we incorporated information from $R(D^{(*)})$ (Sc.1), $B$ decays (Sc.2), and all available data from leptonic and semileptonic decays apart from $R(D^*)$ (Sc.3). We showed that these scenarios can be differentiated by coming data,
using information e.g. from differential decay rates and/or spin and angular asymmetries. These observables therefore allow to verify this hint for NP in semileptonic decays, and gather additional information on its precise nature. Furthermore we pointed out several combinations of observables independent of this kind of NP, as well as common characteristics, which will allow additionally to test for the presence of NP with other Dirac structures. The coming experimental analyses for these modes will therefore be an important step in our quest for NP.

\section*{Acknowledgements}

A. P. would like to thank the Physics Department and the Institute for Advanced Study of the Technical University of Munich for their hospitality during the initial stages of this work, and the support of the Alexander von Humboldt Foundation. This work has been supported in part by the Spanish Government~[grants FPA2007-60323, FPA2011-23778 and CSD2007-00042~(Consolider Project CPAN)]. X.~Q.~L. is also supported in part by the National Natural Science Foundation of China~(NSFC) under contract No.~11005032, the Specialized Research Fund for the Doctoral Program of Higher Education of China~(Grant No.~20104104120001) and the Scientific Research Foundation for the Returned Overseas Chinese Scholars, State Education Ministry. M.~J. is supported by the Bundesministerium f\"ur Bildung und Forschung~(BMBF). The work of A. C. is funded through an FPU grant (AP2010-0308, MINECO, Spain).

\begin{appendix}

\section*{\centering Appendix}

\section{Input parameters and statistical treatment}
\label{subsec:inputs}

Bounds on the parameter space are obtained using the statistical treatment based on frequentist statistics and Rfit for the theoretical uncertainties~\cite{Hocker:2001xe}, which has been implemented in the CKMfitter package~\cite{Charles:2004jd}.
To fix the values of the relevant CKM entries, we only use determinations that are not sensitive to the scalar NP contributions~\cite{Charles:2004jd,Bona:2005vz,Antonelli:2009ws,Amhis:2012bh}. Explicitly, we use the $V_{ud}$ value extracted from super-allowed nuclear $\beta$ decays and the CKM unitarity to determine $V_{us} \equiv \lambda$. The values of $|V_{ub}|$ and $|V_{cb}|$ are determined from exclusive and inclusive $b\to u \ell\bar\nu_{\ell}$ and $b\to c \ell\bar\nu_{\ell}$ transitions, respectively. Relevant hadronic input parameters are collected in Table~\ref{tab::hadronic}, while quark and meson masses as well as any other relevant parameters that do not appear in this table are taken from \cite{Beringer:1900zz}.

%%%%%%%%%%%%%%%%%%%%%%%%%%%%%%%%%%%%%%%%%%%%%%%%%%%%%%%%%%%%%%%%%%%%%%%%%%
\begin{table}[t]
\begin{center}
\caption{\label{tab::hadronic} \it \small Input values for the hadronic parameters, obtained as described in the text. The first error denotes the statistical uncertainty, and the second the systematic/theoretical. ${}^\dagger$This value includes the correction to the isospin limit usually assumed in lattice calculations~\cite{Cirigliano:2011tm,Cirigliano:2011ny}.}
\vspace{0.2cm}
\doublerulesep 0.8pt \tabcolsep 0.5in
\small{
\begin{tabular}{lcc}\hline\hline
Parameter                                            & Value                                   &
Comment \\ \hline
$f_{B_s}$                                            & $(0.228\pm0.001\pm0.006)$~GeV           & \cite{McNeile:2011ng,Bazavov:2011aa,Na:2012kp}\\
$f_{B_s}/f_{B_d}$                                    & $1.198 \pm0.009\pm0.025$                & \cite{Bazavov:2011aa,Na:2012kp,Albertus:2010nm}\\
$f_{D_s}$                                            & $(0.249\pm0.001\pm0.004)$~GeV           & \cite{Bazavov:2011aa,Na:2012kp,Albertus:2010nm,Davies:2010ip}\\
$f_{D_s}/f_{D_d}$                                    & $1.169\pm0.006\pm0.02$                  & \cite{Bazavov:2011aa,Na:2012kp,Albertus:2010nm,Davies:2010ip,Follana:2007uv}\\
$f_K/f_{\pi}$                                        & $1.1908 \pm 0.0016 \pm 0.0104^\dagger$  & \cite{Follana:2007uv,Bernard:2007ps,Durr:2010hr} \\
$\delta_{\mathrm{em}}^{K\ell 2/\pi \ell 2}$          & $-0.0069 \pm 0.0017$                    & \cite{Marciano:2004uf,Cirigliano:2007ga,Cirigliano:2007xi,Antonelli:2010yf,Cirigliano:2011ny} \\
$\delta_{\mathrm{em}}^{\tau K 2/\tau \pi 2}$         & $0.0005 \pm 0.0053$                 & \cite{Marciano:1993sh,Decker:1994ea,Decker:1994kw}
\\ \hline
$\lambda$                                            & $0.2254\pm0.0010$                       & \cite{Hardy:2008gy} \\
$|V_{ub}|$                                           & $(3.51\pm 0.11\pm 0.02) \times 10^{-3}$ &  \cite{Amhis:2012bh} \\
$|V_{cb}|$                                           & $(40.9\pm 1.1)          \times 10^{-3}$ &  \cite{Amhis:2012bh}
\\ \hline
$\rho_1^2$                                           & $1.186 \pm0.036 \pm 0.041$              & \cite{Amhis:2012bh}\\
$G_1(1) |V_{cb}| $                                     & $ (42.64 \pm 1.53) \times 10^{-3} $     & \cite{Amhis:2012bh}\\
$\Delta|_{B\to D l\nu}$                              & $0.46\pm0.02$                           & \cite{Becirevic:2012jf,Bailey:2012jg,deDivitiis:2007uk}
\\\hline
$h_{A_1}(1) |V_{cb}|$                                & $ (35.90 \pm 0.45) \times 10^{-3} $     & \cite{Amhis:2012bh}\\
$R_1(1)$                                             & $1.403 \pm 0.033 $                      & \cite{Amhis:2012bh}\\
$R_2(1)$                                             & $0.854 \pm 0.020 $                      & \cite{Amhis:2012bh}\\
$R_3(1)$                                             & $ 0.97 \pm 0.10 $                       & \cite{HQET-forR31}\\
$\rho^2$                                             & $1.207 \pm0.026 $                       & \cite{Amhis:2012bh}\\
\hline\hline
\end{tabular}}
\end{center}
\end{table}
%%%%%%%%%%%%%%%%%%%%%%%%%%%%%%%%%%%%%%%%%%%%%%%%%%%%%%%%%%%%%%%%%%%%%%%%%%

The plots for the differential observables are obtained using the allowed NP parameter ranges from a different fit. As the latter already include uncertainties from the hadronic input parameters, we do not vary them again additionally.

\section{Kinematics for semileptonic decays}
\label{subsec:kinematics}

Within the SM, the squared matrix element for the decay $\bar B(p_B) \to D^{(*)}(p_{D^{(*)}}, \lambda_{D^{(*)}}) l(k_{l}, \lambda_l) \bar \nu(k_{\bar\nu})$ can be written as~\cite{Korner:1989qb,Hagiwara:1989gza}
\be \label{eq:ff1}
|{\cal{M}}(\bar B \to D^{(*)} l \bar\nu)|^2 = |\langle D^{(*)} l \bar\nu |\cL_{\rm eff}|\bar B\rangle|^2 = L_{\mu \nu} H^{\mu \nu} \,,
\ee
where the leptonic~($L_{\mu \nu}$) and hadronic~($H^{\mu \nu}$) tensors are built from the respective
tensor products of the lepton and hadron currents.
Using the completeness relation for the virtual $W^*$ polarization vectors $\bar \epsilon_{\mu}(\pm,0,t)$, one can further express Eq.~\eqref{eq:ff1} as
\be \label{eq:ff}
|{\cal{M}}(\bar B \to D^{(*)} l \bar \nu)|^2\; = \;\sum_{m, m^{\prime}, n, n^{\prime}} L(m,n) H(m^{\prime},n^{\prime})\, g_{m m ^{\prime}} g_{n n^{\prime}}\,,
\ee
where $g_{m m^{\prime}} = \text{diag}(+1,-1,-1,-1)$, $L(m,n) = L^{\mu \nu}\,\bar \epsilon_{\mu}(m) \bar \epsilon_{\nu}^*(n)$ and $H(m,n) = H^{\mu \nu}\,\bar \epsilon_{\mu}^*(m) \bar \epsilon_{\nu}(n)$.  The two quantities $L(m,n)$ and $H(m,n)$ are Lorentz invariant and can, therefore, be evaluated in different reference frames. For convenience, the hadronic part $H(m,n)$ is usually evaluated in the $B$-meson rest frame with the $z$ axis along the $D^{(*)}$ trajectory, and
$L(m,n)$ in the $l$-$\bar \nu$ center-of-mass frame ({\it i.e.} in the virtual $W^*$ rest frame)~\cite{Korner:1989qb,Hagiwara:1989gza}.

In the $B$-meson rest frame with the $z$ axis along the $D^{(*)}$ trajectory, a suitable basis for the virtual $W^*$ polarization vectors $\bar \epsilon_{\mu}(\pm,0,t)$ can be chosen as~\cite{Korner:1989qb}
\begin{align} \label{eq:basis}
\bar \epsilon_{\mu}(\pm)\, &=\, \frac{1}{\sqrt{2}}\, (0,\pm 1,-i,0) \,, \qquad
\bar \epsilon_{\mu}(  0)\,  =\, \frac{1}{\sqrt{q^2}}\, (|\vec{\textbf{p}}|,0,0,-q_0) \,, \nonumber \\
\bar \epsilon_{\mu}(  t)\, &=\, \frac{1}{\sqrt{q^2}}\, (q_0,0,0,-|\vec{\textbf{p}}|) \,,
\end{align}
where $q_0=(m_B^2 -m_{D^{(*)}}^2 +q^2)/2 m_B$ and $|\vec{\textbf{p}}|=\lambda^{1/2}(m_B^2,m_{D^{(*)}}^2,q^2)/2 m_B$ are the energy and momentum of the virtual $W^*$, with $q^2=(p_B-p_{D^{(*)}})^2$ being the momentum transfer squared, bounded at $m_l^2\leq q^2\leq (m_B -m_{D^{(*)}})^2$, and $\lambda(a,b,c) = a^2 + b^2 +c^2 - 2(a b + b c + c a)$. Similarly, a convenient basis for the $D^*$ polarization vectors is
\begin{align} \label{eq:Dstar-epsilon}
\epsilon_{\alpha}(\pm)\, =\, \mp \frac{1}{\sqrt{2}}\, (0,1,\pm i ,0) \,, \qquad   \epsilon_{\alpha}(0)\, =\, \frac{1}{m_{D^{*}}}\,  (|\vec{\textbf{p}}|, 0,0,E_{D^*}) \,,
\end{align}
where $E_{D^{(*)}} = (m_B^2 + m_{D^{(*)}}^2 - q^2)/2 m_B $ is the $D^{(*)}$ energy in the $B$-meson rest frame.

In the $l$-$\bar \nu$ center-of-mass frame, which can be obtained by a simple boost from the $B$-meson rest frame, the lepton and antineutrino four-momenta are given, respectively, as
\be
k_l=(E_l,p_l \sin\theta,0,p_l \cos\theta)\,,\qquad k_{\bar\nu}=(p_l,-p_l \sin\theta,0,-p_l \cos\theta) \,,
\ee
where $E_l=(q^2+m_l^2)/2 \sqrt{q^2}$, $p_l=(q^2-m_l^2)/2 \sqrt{q^2}$, and $\theta$ is the angle between the $D^{(*)}$ and $l$ three-momenta in this frame. The virtual $W^*$ polarization vectors $\bar \epsilon_{\mu}(\pm,0,t)$ reduce to~\cite{Korner:1989qb,Hagiwara:1989gza}
\begin{align}
\bar \epsilon_{\mu}(\pm)\, &=\, \frac{1}{\sqrt{2}}\, (0,\pm1,-i,0) \,,\qquad
\bar \epsilon_{\mu}(  0)\,  =\,  (0,0,0,-1) \,, \nonumber \\
\bar \epsilon_{\mu}(  t)\, &=\,  \frac{1}{\sqrt{q^2}}\,q_{\mu} \, =\, (1, 0,0, 0) \,.
\end{align}

With the above specified kinematics, the explicit expression for $L_{\mu \nu} H^{\mu \nu}$ in terms of the $q^2$ dependent helicity amplitudes can be found in Refs.~\cite{Korner:1989qb,Fajfer:2012vx}. Using the equations of motion, the hadronic and
leptonic amplitudes of the scalar and the pseudoscalar current can be related to those of the vector and the axial-vector current, respectively. Therefore, the scalar NP contributions can be considered together with the spin-zero component~($\lambda_{W^*} = t$) of the virtual $W^*$ exchange.

\section{Formulae for $\bar B \to D l \bar\nu$ }
\label{subsec:BD}

In the presence of NP of the form \eqref{genlagrangian}, the non-zero hadronic matrix elements of the $\bar B \to D$ transition can be parametrized as
\beqn
\langle D(p_D)|\bar c \gamma^{\mu} b|\bar B(p_B)\rangle & = & f_+(q^2)\, \left[ (p_B + p_D)^{\mu}  - \frac{m_B^2 - m_D^2}{q^2} q^{\mu} \right] + f_0(q^2)\, \frac{m_B^2 - m_D^2}{q^2} q^{\mu} \,, \\
\langle D(p_D)|\bar c\, b|\bar B(p_B)\rangle & = & \frac{q_{\mu}}{\overline{m}_b - \overline{m}_c}\; \langle D(p_D)|\bar c \gamma^{\mu} b|\bar B(p_B)\rangle\; =\; \frac{m_B^2 - m_D^2}{\overline{m}_b - \overline{m}_c}\, f_0(q^2)\,,
\eeqn
where $\overline{m}_q$ are the running quark masses and the two QCD form factors $f_+(q^2)$ and $f_0(q^2)$ encode the strong-interaction dynamics. Contracting the above matrix elements with the virtual $W^*$ polarization vectors \eqref{eq:basis} in the $B$-meson rest frame, we obtain the two non-vanishing helicity amplitudes~\cite{Korner:1989qb,Fajfer:2012vx}:
\beqn \label{eq:BD-helicities}
H_{0}(q^2) &=& \frac{2 m_B |\vec{\textbf{p}}|}{\sqrt{q^2}}\, f_+(q^2) \,, \nonumber \\
H_{t}(q^2) &=& \frac{m_B^2 - m_D^2}{\sqrt{q^2}}\, f_0(q^2) \left[1 + \delta_{cb}^l\,
\frac{q^2}{(m_B-m_D)^2} \right] \,,
\eeqn
where $\delta_{cb}^l$, defined by Eq.~\eqref{eq:ww}, accounts for the contribution from the charged scalar.

It is customary to relate the QCD form factors $f_+(q^2)$ and $f_0(q^2)$ to the quantities $G_1(w)$ and $\Delta(w)$ in the HQET~\cite{HQET-forR31}
\be
f_+(q^2)\, = \, \frac{G_1(w)}{R_D} \,, \qquad
f_0(q^2)\, = \, R_D\, \frac{(1 + w)}{2}\, G_1(w)\, \frac{1+r}{1-r}\, \Delta(w) \,,
\ee
where $R_{D^{(*)}} = 2 \sqrt{m_{B} m_{D^{(*)}}}/(m_{B} + m_{D^{(*)}})$, $r=m_{D^{(*)}}/m_{B}$, and the new kinematical variable $w$ is defined as
\begin{equation} \label{def:wR}
w \,=\, v_B \cdot v_{D^{(*)}}\, = \, \frac{m_{B}^2 + m_{D^{(*)}}^2 - q^2}{2 m_{B} m_{D^{(*)}}} \,,
\end{equation}
with $v_B$ and $v_{D^{(*)}}$ being the four-velocities of the $B$ and $D^{(*)}$ mesons, respectively. We approximate the scalar density $\Delta(w)$ by a constant value $\Delta(w) = 0.46 \pm 0.02$~\cite{Becirevic:2012jf,Bailey:2012jg,deDivitiis:2007uk} and $G(w)$ is parametrized in terms of the normalization $G_1(1)$ and the slope $\rho_1^2$ as~\cite{Caprini:1997mu}
\be
G_1(w)\, = \, G_1(1)\, \left[1- 8 \rho_1^2\, z(w) + (51 \rho_1^2 -10)\, z(w)^2 - (252 \rho_1^2 - 84)\, z(w)^3   \right] \,,
\ee
with $z(w)=(\sqrt{w+1}-\sqrt{2})/(\sqrt{w+1}+\sqrt{2})$.

Equipped with the above information, the double differential decay rates for $\bar B \to D l \bar\nu$, with $l$ in a given helicity state~($\lambda_l=\pm 1/2$), can be written as
\beqn \label{eq:double-B2D}
\frac{d^2 \Gamma^D[\lambda_{l} = -1/2]}{dq^2 d \cos\theta} &=&  \dfrac{G_F^2 |V_{cb}|^2 q^2}{128 \pi^3 m_B^2}\, \left( 1 - \frac{m_{l}^2}{q^2} \right)^2\, |\vec{\textbf{p}}|\, |H_0(q^2)|^2\, \sin^2\theta \,,  \nonumber \\[0.2cm]
\frac{d^2 \Gamma^D[\lambda_{l} = +1/2]}{dq^2 d \cos\theta} &=&  \dfrac{G_F^2 |V_{cb}|^2 q^2}{128 \pi^3 m_B^2}\, \left( 1 - \frac{m_{l}^2}{q^2} \right)^2\, |\vec{\textbf{p}}|\; \frac{m_{l}^2}{q^2}\, |H_0(q^2) \cos\theta - H_{t}(q^2)|^2 \,,
\eeqn
from which the total decay rate and the various $q^2$-dependent observables can be obtained via summation over $\lambda_l$ and/or integration over $\cos\theta$. Owing to its lepton-mass suppression, the $\lambda_{l} = +1/2$ helicity amplitude is only relevant for the $\tau$ decay mode.

\section{Formulae for $\bar B \to D^* l \bar\nu$}
\label{subsec:BDstar}

For the $\bar B \to D^*$ transition, the hadronic matrix elements of the vector and axial-vector currents are described by four QCD form factors $V(q^2)$ and $A_{0,1,2}(q^2)$ via
\begin{align} \label{eq:matrix}
\langle D^*(p_{D^*},\epsilon^*) |  \bar c \gamma_{\mu} b | \bar B(p_B) \rangle \; &=\; \frac{2 i V(q^2)}{m_B + m_{D^*}}\, \epsilon_{\mu \nu \alpha \beta}\, \epsilon^{*\nu} p_B^{\alpha} p_{D^*}^{\beta} \,, \nonumber \\[0.2cm]
\langle D^*(p_{D^*},\epsilon^*) |  \bar c \gamma_{\mu} \gamma_5 b | \bar B(p_B) \rangle \; &=\; 2 m_{D^*}\, A_0(q^2)\, \frac{\epsilon^* \cdot q}{q^2}\, q_{\mu} + (m_B + m_{D^*})\, A_1(q^2)\, \left(\epsilon_{\mu}^* - \frac{\epsilon^* \cdot q}{q^2}\, q_{\mu} \right) \nonumber \\[0.2cm]
&\; - A_2(q^2)\, \frac{\epsilon^* \cdot q}{m_B + m_{D^*}}\, \left[(p_B + p_{D^*})_{\mu} - \frac{m_B^2 - m_{D^*}^2}{q^2}\, q_{\mu} \right]\,,
\end{align}
from which one can show that, using the equations of motion, the $\bar B \to D^*$ matrix element for the scalar current vanishes while the pseudoscalar one reduces to
\begin{align}
\langle D^*(p_{D^*},\epsilon^*) |  \bar c \gamma_{5} b | \bar B(p_B) \rangle\; &=\; -\, \frac{q_{\mu}}{\overline{m}_b +\overline{m}_c}\; \langle D^*(p_{D^*},\epsilon^*) | \bar c \gamma^{\mu} \gamma_{5} b | \bar B(p_B) \rangle \nonumber \\
&=\; -\, \frac{2 m_{D^*}}{\overline{m}_b +\overline{m}_c}\; A_0(q^2)\, \epsilon^* \cdot q.
\end{align}
Contracting the above matrix elements with the $W^*$ and $D^*$ polarization vectors in Eqs.~\eqref{eq:basis} and \eqref{eq:Dstar-epsilon}, we obtain the four non-vanishing helicity amplitudes~\cite{Korner:1989qb,Fajfer:2012vx}:
\beqn \label{eq:BDstar-helicities}
H_{\pm \pm}(q^2) &=& (m_B + m_{D^*})\, A_1(q^2) \mp \frac{2 m_B}{m_B + m_{D^*}}\, |\vec{\textbf{p}}|\, V(q^2) \,, \nonumber \\[0.2cm]
H_{00}(q^2) &=& \frac{1}{2 m_{D^*} \sqrt{q^2}}\, \left[(m_B^2 - m_{D^*}^2 -q^2)\,(m_B + m_{D^*})\, A_1(q^2) - \frac{4 m_B^2 |\vec{\textbf{p}}|^2}{m_B + m_{D^*}}\, A_2(q^2) \right]\,, \nonumber \\[0.2cm]
H_{0t}(q^2) &= & \frac{2 m_B |\vec{\textbf{p}}|}{\sqrt{q^2}}\, A_0(q^2)\, \left(1 - \Delta_{cb}^l \, \frac{q^2}{m_B^2}\right)\,,
\eeqn
where $\Delta_{cb}^l$, defined by Eq.~\eqref{eq::Delta}, accounts for the contribution from the charged scalar.

In the heavy-quark limit for the $b$ and $c$ quarks, the four QCD form factors $V(q^2)$ and $A_{0,1,2}(q^2)$ are related to the universal HQET form factor $h_{A_1}(w)$ via~\cite{Caprini:1997mu}
\begin{align}
V(q^2)  \; &=\; \frac{R_1(w)}{R_{D^*}}\, h_{A_1}(w) \,,\nonumber \\
A_0(q^2)\; &=\; \frac{R_0(w)}{R_{D^*}}\, h_{A_1}(w) \,, \nonumber \\
A_1(q^2)\; &=\; R_{D^*}\, \frac{w +1}{2}\, h_{A_1}(w) \,, \nonumber \\
A_2(q^2)\; &=\; \frac{R_2(w)}{R_{D^*}}\, h_{A_1}(w) \,,
\end{align}
where the $w$ dependence of $h_{A_1}(w)$ and the three ratios $R_{0,1,2}(w)$ reads~\cite{Caprini:1997mu}
\begin{align}
h_{A_1}(w)\; &=\; h_{A_1}(1)\,[1 - 8 \rho^2 z(w) + (53 \rho^2 - 15)\, z(w)^2 -(231 \rho^2 - 91)\, z(w)^3] \,, \nonumber \\
R_0(w)\; &=\; R_0(1) - 0.11(w-1) + 0.01(w-1)^2 \,, \nonumber \\
R_1(w)\; &=\; R_1(1) - 0.12(w-1) + 0.05(w-1)^2 \,, \nonumber \\
R_2(w)\; &=\; R_2(1) - 0.11(w-1) - 0.06(w-1)^2 \,.
\end{align}
The free parameters $\rho^2$, $R_1(1)$ and $R_2(1)$ are determined from the well-measured $\bar B \to D^* \ell  \bar\nu$ decay distributions~\cite{Amhis:2012bh} ($\ell=e,\mu$), whereas for the parameter $R_0(1)$, that appears only in the helicity-suppressed amplitude $H_{0t}$, we have to rely on the HQET prediction for the linear combination~\cite{HQET-forR31},
\begin{equation}
R_3(1)\; =\; \frac{R_2(1) (1-r) + r\, [ R_0(1) (1 + r) - 2 ]}{(1-r)^2} \; =\; 0.97 \pm 0.10\,,
 \end{equation}
which includes the leading-order perturbative~(in $\alpha_s$) and power~($1/m_{b,c}$) corrections to the heavy-quark limit, plus a conservative $10\%$ uncertainty  to account for higher-order contributions~\cite{Fajfer:2012vx}.

Finally, the double differential decay rates for $\bar B \to D^* l \bar\nu$, with $l$ in a given helicity state~($\lambda_l=\pm 1/2$), can be written as
\beqn \label{eq:double-B2Dstar}
\dfrac{d^2 \Gamma^{D^*} [\lambda_{l} = -1/2]}{dq^2 d\cos\theta} &= & \dfrac{G_F^2 |V_{cb}|^2 |\vec{\textbf{p}}| q^2}{256 \pi^3 m_B^2}\, \left( 1- \frac{m_{l}^2}{q^2}\right)^2  \nonumber \\
&& \times  \left[ (1- \cos\theta)^2\, |H_{++}|^2 + (1 + \cos\theta)^2\, |H_{--}|^2 + 2 \sin^2\theta\, |H_{00}|^2 \right] \,, \nonumber \\[0.2cm]
\dfrac{d^2 \Gamma^{D^*} [\lambda_{l} = +1/2]}{dq^2 d\cos\theta} &= & \dfrac{G_F^2 |V_{cb}|^2 |\vec{\textbf{p}}| q^2}{256 \pi^3 m_B^2}\, \left( 1- \frac{m_{l}^2}{q^2}\right)^2\, \frac{m_{l}^2}{q^2}  \nonumber \\
&& \times \left[ \sin^2\theta\, (|H_{++}|^2 + |H_{--}|^2) + 2\, |H_{0t} - H_{00} \cos\theta |^2 \right] \,,
\eeqn
which are the starting point for the total decay rate, as well as the additional observables considered.

\end{appendix}

\end{document}